\documentclass[a4paper,11pt]{article}
\usepackage{cite}
\usepackage[all]{xy}
\usepackage{amsmath,amsfonts,amssymb,amsthm,epsfig,amscd,comment,latexsym,psfrag,booktabs}
\usepackage{CJK}
\usepackage{mathrsfs}
\usepackage{nicefrac,xspace,tikz}
\usepackage{arydshln}
\usepackage{stmaryrd}
\usepackage{graphicx,bm}
\usepackage{amscd}
\usepackage{multirow}
\usepackage{makecell}
\usetikzlibrary{arrows}

\numberwithin{equation}{section}

\makeatletter

\newcommand{\Rmnum}[1]{\expandafter\@slowromancap\romannumeral #1@}
\makeatother
\usepackage{graphicx,amssymb,amsmath,bm,latexsym}
\allowdisplaybreaks

\begin{document}

\begin{titlepage}
\begin{center}
{\Large\bf Calogero-Sutherland-type quantum systems, generalized hypergeometric functions and superintegrability for integral chains}
\vskip .2in
{\large
Fan Liu$^{a,}$\footnote{liufan-math@cnu.edu.cn},
Rui Wang$^{b,}$\footnote{wangrui@cumtb.edu.cn},
Jie Yang$^{a,}$\footnote{yangjie@cnu.edu.cn}
and
Wei-Zhong Zhao$^{a,}$\footnote{Corresponding author: zhaowz@cnu.edu.cn}} \vskip .2in
$^a${\em School of Mathematical Sciences, Capital Normal University,
Beijing 100048, China} \\
$^b${\em Department of Mathematics, China University of Mining and Technology, Beijing 100083, China}\\
			
\begin{abstract}
We reinvestigate the Calogero-Sutherland-type (CS-type) models and generalized hypergeometric functions. We construct the generalized CS operators for  circular, Hermite, Laguerre, Jacobi and Bessel
cases and establish the generalized Lassalle-Nekrasov correspondence. A family of operators are constructed based on the spherical degenerate double affine Hecke algebra. In terms of these operators,
we provide concise representations and constraints for the generalized hypergeometric functions. We analyze the superintegrability for the $\beta$-deformed integrals, where the measures are associated
with the corresponding ground state wave functions of Hermite, Laguerre, Jacobi and Bessel type CS models. Then based on the generalized Laplace transformation of Jack polynomials, we construct certain
two integral chains and analyze the superintegrability property.

\end{abstract}
\end{center}
		
{\small Keywords: Integrable Hierarchies, Matrix Models}
		
\end{titlepage}

\section{Introduction}
The Calogero-Sutherland (CS) models describe quantum mechanical systems of $N$ particles in one dimension with long-range pairwise interaction (rational potential \cite{Calogero} and trigonometric
potential \cite{Sutherland}, which are called the rational and circular types, respectively). Due to rich integrability structure and widespread applications, the CS models have been generalized in
many aspects, such as the potential associated with Weierstrass elliptic function \cite{Langmann}, the root systems for classical reflection groups beyond the $A_{N-1}$ type \cite{Olshanetsky},
the particles in relativistic quantum field \cite{Ruijsenaars86} or equipped with a $\mathfrak{su}(p)$ spin \cite{Bernard} and two types of particles \cite{Sergeev04,Sergeev09}.  Their integrability
and exact solvability have been extensively studied \cite{Langmann,Olshanetsky,Ruijsenaars86,Ruijsenaars87,Bernard,Sergeev04,Sergeev09,Lapointe,BakerCMP,Forrester1,Forrester2,Forrester3,ForresterLog,CMS,Hallnas09,Hallnas10}.
For the CS-type models, such as rational [1], circular [2], Hermite, Laguerre, Jacobi and Bessel types \cite{BakerCMP, Hallnas09, Hallnas10}, one may obtain the CS operators by the specific transformations
for the Hamiltonians. From an algebraic perspective, these CS operators can be embedded in Yangian of $\mathfrak{gl}_N$ \cite{Bernard} or the spherical degenerate double affine Hecke
algebra ($\mathbf{SH}_N$) \cite{Schiffmann}. These eigenfunctions of CS operators are orthogonal with respect to certain measures which are associated with the corresponding ground wave functions and
have close relationships with the generalized hypergeometric functions \cite{BakerCMP,Macdonald13,Hallnas09,Hallnas10,Yan,Kaneko,Koranyi}.

Much interest has been attributed to matrix models, since they provide a rich set of approaches to physical systems. The relationship between CS models and matrix integrals has attracted considerable
attention \cite{Hikami,zhangwang,Forrester1,Forrester2,Forrester3,ForresterLog,Lesage,Awata95,Awata96}. There is a correspondence between the classical Calogero model and Hermitian one-matrix model \cite{Hikami,zhangwang}.
The Selberg-Aomoto integral can be used to calculate the correlation functions of the CS models \cite{Forrester1,Forrester2,Forrester3,Lesage}, it is also related to the CS models by the integral representation of the Jack polynomials \cite{Awata95,Awata96}. Derived from the Virasoro (or $\widetilde{W}$) constraints, $W$-representations of matrix models realize the partition functions by acting on elementary functions with exponents of
the given $\hat{W}$-operators, which give the dual expressions for partition functions through differentiation rather than integration  \cite{Morozov09,Cassia,MirGKM,Drachov}.
The superintegrability for matrix models \cite{MirSum} can be analyzed conveniently by means of $W$-representations. Here the superintegrability means that for the matrix models, the average of a properly
chosen symmetric function is proportional to ratios of symmetric functions on a proper locus, i.e., $\langle character\rangle \sim character$.
Recently, a considerable effort has been devoted to a family of $\beta$-deformed partition functions with $W$-representations and superintegrability \cite{Rui22,MirSum,MirSkew,MirInter,Oreshina03,Oreshina04}.
Some integral forms \cite{Oreshina03,Oreshina04} have been constructed by the $\beta$-deformed Harish-Chandra-Itzykson-Zuber (HCIZ) integral \cite{Eynard}.
The $\hat{W}$-operators for these partition functions are commutative and can be regarded as the Hamiltonians of certain many-body models \cite{MirMany,MirCom}.
When taking $\beta=1$, these partition functions are KP/Toda $\tau$-functions \cite{MirSkew,MirInter}, since the corresponding $\hat{W}$-operators belong to the algebra $\hat{\mathfrak{gl}}_{\infty}$.
In addition, some of these partition functions are identical to certain weighted Hurwitz $\tau$-functions on orientable ($\beta=1$) \cite{Alexandrov14,Alexandrov24}
or non-orientable ($\beta\neq1$) \cite{Chapuy1,Chapuy2} Riemann surface.

In this paper, we reinvestigate the CS-type models and generalized hypergeometric functions. We shall present the new properties of CS operators and generalized hypergeometric functions.
The goal of this paper is to reveal the relations between the CS-type models, $\beta$-deformed matrix integrals and generalized hypergeometric functions.
The integral chain and its skew version will be constructed, where the former is equal to the generalized hypergeometric function.

This paper is organized as follows.
In section \ref{CSmodel}, we give a brief review for six CS-type models. We analyze the properties of the CS operators and their eigenfunctions. We construct generalized CS operators for the circular,
Hermite, Laguerre, Jacobi and Bessel cases, and establish the generalized Lassalle-Nekrasov correspondence.
In section \ref{WOoperator}, based on the $\mathbf{SH}_N$ algebra, we construct the $\hat{W}$-operators and $\hat O$-operator. Two families of commutative operators are presented, which can be regarded as the
Hamiltonians of certain many-body integrable system.
In section \ref{Hyper}, we discuss the generalized hypergeometric functions and present their $O$-representations. In terms of $\hat{W}$-operators, the constraints for generalized hypergeometric functions
are given. The properties associated with the generalized hypergeometric functions and Hermite and Laguerre polynomials are analyzed.
In section \ref{super}, we reveal the relations between CS-type systems, $\beta$-ensembles and generalized hypergeometric functions. Two integral chains are constructed.
We end this paper with the conclusions in section \ref{Conclusion}.

\section{CS-type quantum systems} \label{CSmodel}	
\subsection{CS-operators}

Let us recall a family of CS-type quantum systems. We start with the Schr\"odinger equation
\begin{eqnarray}
H(\mathbf{x})\Psi_0(\mathbf{x})=E_0\Psi_0(\mathbf{x}),
\end{eqnarray}
where the Hamiltonian is given by
\begin{eqnarray}\label{H}
H(\mathbf{x})=-\sum_{j=1}^{N}\frac{\partial^2} {\partial x_j^2}+\sum_{j=1}^{N}V_1(x_j)+ 2\beta(\beta-1)\sum_{1\le j<k\le N}V_2(x_j,x_k),
\end{eqnarray}
$\mathbf{x}=(x_1,\cdots,x_N)$, $\beta\in\mathbb{R}_+$, $V_1(x)$ and $V_2(x,y)$ are real-valued functions,
and the ground state is factorized into the form
\begin{eqnarray}\label{Psi0}
\Psi_0(\mathbf{x})=\prod_{i=1}^{N}f_1(x_i)\prod_{1\le j<k\le N}(f_2(x_j,x_k))^{\beta}.
\end{eqnarray}

We may derive the compatibility conditions between the potentials $V_1(x)$, $V_2(x,y)$ and ground state
\begin{eqnarray}\label{V1f1}
	&&V_1(x)-\frac{1}{f_1(x)}\left(\frac{\partial^2}{\partial x^2}f_1(x)\right)=c_1,\nonumber\\
	&&V_2(x,y)-\frac{1}{2(f_2(x,y))^{2}}\left((\frac{\partial}{\partial x}f_2(x,y))^2+(\frac{\partial} {\partial y}f_2(x,y))^2\right)=c_2,\nonumber\\
    &&\frac{\frac{\partial}{\partial x}f_1(x)}{f_1(x)}\frac{\frac{\partial}{\partial x} f_2(x,y)}{f_2(x,y)}+
    \frac{\frac{\partial}{\partial y}f_1(y)}{f_1(y)}\frac{\frac{\partial}{\partial y} f_2(x,y)}{f_2(x,y)}
	+\frac{1}{2}\frac{(\frac{\partial^2}{\partial x^2}+\frac{\partial^2}{\partial y^2})f_2(x,y)}{f_2(x,y)}=c_3,\nonumber\\
	&&\sum_{i=1}^{N}\sum_{i\neq j\neq k}\frac{\frac{\partial}{\partial x_i}f_2(x_i,x_j)}{f_2(x_i,x_j)}\frac{\frac{\partial}{\partial x_i}f_2(x_i,x_k)}{f_2(x_i,x_k)}=c_4,
\end{eqnarray}
where $c_k$, $k=1,2,3,4$, are constants and the ground state energy is given by
\begin{eqnarray}
E_0=Nc_1+N(N-1)\beta[(\beta-1)c_2-c_3]-\beta^2c_4.
\end{eqnarray}

The equations (\ref{V1f1}) provide the exact relationship between the Hamiltonian (\ref{H}) and the ground-state wave function (\ref{Psi0}).
Once the Hamiltonian is given, the ground state wave function can be determined, and vice versa.
We list six CS-type systems in the literature in Table \ref{Tab1}.

\begin{table*}[h!]
	\small
	\renewcommand\arraystretch{2.5}
	\centering
	\caption{Six CS-type models, the types $R$, $C$, $H$, $L$, $J$ and $B$ stand for rational, circular, Hermite, Laguerre, Jacobi and Bessel, respectively, and the last four are named by the corresponding orthogonal polynomials.}\label{Tab1}
	\begin{tabular}{c|c|c|c|c|c}
		\toprule[2pt]
		\multirow{2}*{$X$}&\multicolumn{2}{c|}{Hamiltonian $H^{(X)}$}&
		\multicolumn{2}{c|}{Ground state $\Psi_0^{(X)}$}& \multirow{2}*{\makecell{Root system\\and  potential}}\\
		\cline{2-5} ~\multirow{2}*{}
		&$V_1(x)$ & $V_2(x,y)$ & $f_1(x)$& $f_2(x,y)$ & \\
		\midrule
		$R$ & 0 & $(x-y)^{-2}$ & $1$ & $x-y$ &\makecell{$A_{N-1}$\\ rational} \\
		$C$ & 0 & $ ({\pi\over L})^2\sin^{-2}({\pi\over L}(x-y))$ & $1$ & $\sin({\pi\over L}(x-y))$ &\makecell{$A_{N-1}$\\ trigonometric} \\
		$H$ & $\omega^2 x^2$ & $(x-y)^{-2}$ & $e^{-\frac{\omega}{2}x^2}$ &$x-y$& \makecell{$A_{N-1}$ rational\\harmonic} \\
		$L$ & $(a^2-{1\over4})x^{-2}+\omega^2x^2$ & $(x-y)^{-2}+(x+y)^{-2}$ &$x^{a+{1\over2}}e^{-\frac{\omega}{2}x^2}$ & $x^2-y^2$ & \makecell{$B_N$ rational\\harmonic}   \\
		$J$ & \makecell{$(a^2-{1\over4}) \sin^{-2}x$\\$+(b^2-{1\over4})\cos^{-2}x$} & \makecell{$\sin^{-2}(x-y)$\\$+\sin^{-2}(x+y)$} & \makecell{$(\sin x)^{a+{1\over2}}$\\$\times(\cos x)^{b+{1\over2}}$} & \makecell{$\sin^2 x$\\$-\sin^2 y$} & \makecell{$BC_N$\\ trigonometric}\\
		$B$ & $e^{-2x}+ae^{-x}$ & $\frac{1}{4}\sinh^{-2}[(x-y)/2]$ & $e^{\frac{a+1}{2}x-e^{-x}}$ & $e^x-e^y$ & \makecell{$A_{N-1}$\\ hyperbolic} \\
		\bottomrule[2pt]
	\end{tabular}
\end{table*}
The energies for corresponding ground states are given by
\begin{subequations}
\begin{align}
	&E_0^{(R)} = 0,\\
	&E_0^{(C)} =\frac{\beta^{2}}{3}\left({\pi\over L}\right)^2 N(N^{2}-1), \\
	&E_0^{(H)} =\omega N[\beta(N-1)+1], \\
	&E_0^{(L)} =\omega N[2\beta(N-1)+2a+2], \\
	&E_0^{(J)} =\beta^2N[(N-1)(4(N-2)+a+b+1) +(a+b+1)^2]+2\beta N(N-1), \\
	&E_0^{(B)} =-\frac{\beta^2}3N(N^2-1) -\frac{\beta(a+1)-\beta^2}2N(N-1)-\frac{(a+1)^2}4N.
\end{align}
\end{subequations}

The models $H^{(R)}$ and $H^{(C)}$ were first proposed by Calogero \cite{Calogero} and Sutherland \cite{Sutherland}, respectively.
The models $H^{(H)}$, $H^{(L)}$,  $H^{(J)}$ \cite{BakerCMP,ForresterLog} and $H^{(B)}$ \cite{Hallnas09} were analyzed well.
All of these models are related to root systems for classical reflection groups \cite{Olshanetsky}.

By the transformation
\begin{eqnarray}
\mathcal{H}^{(X)}=-\eta^{(X)}(\Psi_0^{(X)})^{-1}\left(H^{(X)}-E_0^{(X)}\right)\Psi_0^{(X)},
\end{eqnarray}
where $X\in\{R,C,H,L,J,B\}$, the normalized coefficients are $\eta^{(H)}=1/\omega$, $\eta^{(L)}=1/4\omega$
and $\eta^{(R)}=\eta^{(C)}=\eta^{(J)}=\eta^{(B)}=1$,
we can obtain the monic CS operators (after some coordinate transformations)
\begin{subequations}\label{CSoperator}
\begin{align}
	\mathcal{H}^{(R)}(\mathbf{y})&=\sum_{j=1}^{N} \frac{\partial^2}{\partial y_j^2}+ 2\beta\sum_{1\leq j\neq k\leq N} \frac{1}{y_j-y_k} \frac{\partial}{\partial y_j}, & &(y_j=x_j),\\
	\mathcal{H}^{(C)}(\mathbf{y})&=\sum_{j=1}^{N}\left(y^2_j \frac{\partial^2}{\partial y_j^2}-[\beta(N-1)-1]y_j \frac{\partial}{\partial y_j}\right)& &\nonumber\\
	&+2\beta\sum_{1\leq j\neq k\leq N} \frac{y_j^2}{y_j-y_k} \frac{\partial}{\partial y_j}, & &(y_j=e^{{2\pi\mathbf{i}\over L}x_j}),\\
	\mathcal{H}^{(H)}(\mathbf{y})&=\sum_{j=1}^{N}\left(\frac{\partial^2}{\partial y_j^2}-2y_j \frac{\partial}{\partial y_j}\right)+ 2\beta\sum_{1\leq j\neq k\leq N} \frac{1}{y_j-y_k} \frac{\partial}{\partial y_j},& &(y_j=\sqrt{\omega}x_j),\\
	\mathcal{H}^{(L)}(\mathbf{y})&=\sum_{j=1}^{N}\left(y_j \frac{\partial^2}{\partial y_j^2}+ (a+1- y_j) \frac{\partial}{\partial y_j}\right)+2\beta\sum_{1\leq j\neq k\leq N}\frac{y_j}{y_j-y_k} \frac{\partial}{\partial y_j},& &(y_j=\omega x_j^2),\\
	\mathcal{H}^{(J)}(\mathbf{y})&= \sum_{j=1}^{N}\left(y_j(1-y_j) \frac{\partial^2}{\partial y_j^2}+[a+1-(a+b+2)y_j] \frac{\partial}{\partial y_j}\right)& &\nonumber \\
	&+2\beta\sum_{1\leq j\neq k\leq N} \frac{y_j(1-y_j)}{y_j-y_k} \frac{\partial}{\partial y_j},& &(y_j=\sin^2 x_j),\\
	\mathcal{H}^{(B)}(\mathbf{y})&=\sum_{j=1}^{N}\left(y^2_j\frac{\partial^2}{\partial y_j^2}+[2+(a+2)y_j] \frac{\partial}{\partial y_j}\right)+2\beta\sum_{1\leq j\neq k\leq N}\frac{y_j^2}{y_j-y_k}\frac{\partial}{\partial y_j}, & &(y_j=e^{x_j}),
\end{align}
\end{subequations}
where $\mathbf{y}=(y_1,\cdots,y_N)$.

\subsection{Generalized Lassalle-Nekrasov correspondence}
The CS operators (\ref{CSoperator}) can be rewritten as
\begin{eqnarray}\label{dep}
\mathcal{H}^{(X)}=\mathcal{H}_0^{(X)}+\mathcal{H}_{<0}^{(X)},
\end{eqnarray}
where $X\in\{R,C,H,L,J,B\}$, $\mathcal{H}_0^{(X)}$ and $\mathcal{H}_{<0}^{(X)}$ are homogeneous differential operators and the degrees satisfy $\deg(\mathcal{H}_0^{(X)})=0$ and  $\deg(\mathcal{H}_{<0}^{(X)})<0$.
Their explicit expressions are listed in Table \ref{Tab2}. For convenience, we introduce the operators $\hat{E}_r$ and $\hat{D}_s$ in Table \ref{Tab2} as follows:
\begin{eqnarray}
&&\hat{E}_r=\sum_{j=1}^{N}x_j^r\frac{\partial}{\partial x_j},\label{Er}\nonumber\\
&&\hat{D}_s=\sum_{j=1}^{N}x_j^s\frac{\partial^2}{\partial x_j^2}+2\beta\sum_{j\neq k}\frac{x_j^s}{x_j-x_k}\frac{\partial}{\partial x_j},
\end{eqnarray}
where $r=0,1$, $s=0,1,2$ and $\deg(\hat{E}_{r+1})=\deg(\hat{D}_{r+2})=r$.

Let us consider the Jack polynomial with normalized coefficients \cite{Kaneko}
\begin{eqnarray}\label{nJack}
P_{\lambda}^{(C)}=|\lambda|! \beta^{-|\lambda|}j^{-1}_{\lambda}J_{\lambda},
\end{eqnarray}
where $J_{\lambda}$ is the integral form of Jack polynomial associated with the partition $\lambda=(\lambda_1,\cdots,\lambda_{l(\lambda)})$ \cite{Stanley89,Macdonald}, $|\lambda|=\sum_{i=1}^{l(\lambda)}\lambda_i$, $(i,j)\in\lambda$ iff $1\le i\le l(\lambda)$ and $1\le j\le \lambda_i$, $j_{\lambda}=\prod_{(i,j)\in\lambda}h^*_\lambda(i,j)h_*^\lambda(i,j)$ is the normalization factor, $h^*_\lambda(i,j)=(\lambda^T_j-i)+\beta^{-1}(\lambda_i-j+1)$ and $h_*^\lambda(i,j)=h^*_\lambda(i,j)+1-\beta^{-1}$.

The circular CS operator $\mathcal{H}^{(C)}$ satisfies
\begin{eqnarray}
\mathcal{H}^{(C)}P^{(C)}_{\lambda}=\epsilon^{(C)}_{\lambda}P^{(C)}_{\lambda},
\end{eqnarray}
where $\epsilon^{(C)}_{\lambda}=\sum_{(i,j)\in\lambda} 2j-1+\beta(N-2i+1)$.

In terms of the Jack polynomial (\ref{nJack}), one can define the non-homogeneous symmetric polynomials by the following two conditions \cite{Macdonald,Diejen}
\begin{subequations}
\begin{eqnarray}\label{defPX}
	&&P^{(X)}_{\lambda}=c_{\lambda,\lambda}^{(X)} P^{(C)}_{\lambda}+\sum_{\mu\subset\lambda}c_{\lambda,\mu}^{(X)} P^{(C)}_{\mu},\label{uptra}\\
	&&\mathcal{H}^{(X)}P^{(X)}_{\lambda}=\epsilon^{(X)}_{\lambda}P^{(X)}_{\lambda}, \label{eigenPX}
	\end{eqnarray}
\end{subequations}
where $X\in\{H,L,J,B\}$, $c^{(X)}_{\lambda,\mu}\in\mathbb{C}$ are the elements of specific invertible constant matrices and the eigenvalues $\epsilon_{\lambda}^{(X)}$ are defined by solving
\begin{eqnarray}\label{eigenPC}
\mathcal{H}^{(X)}_0 P^{(C)}_{\lambda} =\epsilon^{(X)}_{\lambda}P^{(C)}_{\lambda}.
\end{eqnarray}
The explicit values of $\epsilon^{(X)}_{\lambda}$ are determined by following formulas \cite{Stanley89,Macdonald}
\begin{eqnarray}
\hat{E}_1 P^{(C)}_{\lambda} &=&|\lambda| P^{(C)}_{\lambda},\nonumber\\
\hat{D}_2 P^{(C)}_{\lambda} &=&2\sum_{(i,j)\in\lambda}[j-1+\beta(N-i)] P^{(C)}_{\lambda}.
\end{eqnarray}

We list the eigenvalues $\epsilon^{(X)}_{\lambda}$ and symmetric polynomials $P^{(X)}_{\lambda}$ for the CS operators $\mathcal{H}^{(X)}$ in Table \ref{Tab2}.
Note that the symmetric polynomials vanish for the rational case due to $\mathcal{H}_0^{(R)}=0$.

Taking $c_{\lambda,\lambda}^{(H)}= \frac{2^{|\lambda|}}{P_{\lambda}^{(C)}(1^N)}$, $c_{\lambda,\lambda}^{(L)}= \frac{(-1)^{|\lambda|}}{|\lambda|!P_{\lambda}^{(C)}(1^N)}$
and  $c_{\lambda,\lambda}^{(J)}= c_{\lambda,\lambda}^{(B)}=1$
in (\ref{uptra}), we see that $P^{(H)}_{\lambda}$, $P^{(L,a)}_{\lambda}$, $P^{(J,a,b)}_{\lambda}$ and $P^{(B,a)}_{\lambda}$
become the $\beta$-deformed multivariate Hermite, Laguerre, Jacobi \cite{Lassalle1,Lassalle2,Lassalle3} and Bessel polynomials \cite{Hallnas09}, respectively.

\begin{table*}[h!]
\renewcommand\arraystretch{2.5}
\centering
\caption{The CS operators $\mathcal{H}^{(X)}_{0}$, $\mathcal{H}^{(X)}_{<0}$, eigenvalues $\epsilon^{(X)}_{\lambda}$ and symmetric polynomials $P^{(X)}_{\lambda}$}\label{Tab2}
\begin{tabular}{c|c|c|c|c}
	\toprule[2pt]
	&  $\mathcal{H}^{(X)}_{0}$ & $\mathcal{H}^{(X)}_{<0}$ & $\epsilon^{(X)}_\lambda$ & $P^{(X)}_{\lambda}$ \\
	\midrule
	$R$ &  $0$ & $\hat{D}_0$ & - & - \\
	$C$ &  $\hat{D}_2+(1+\beta-N\beta)\hat{E}_1$ & 0 & $\sum_{(i,j)\in\lambda}2j-1+\beta(N-2i+1)$ & \makecell{$P^{(C)}_{\lambda}$\\(Jack)}  \\
	$H$ &   $-2\hat{E}_1$ & $\hat{D}_0$ &  $-2|\lambda|$ & \makecell{$P^{(H)}_{\lambda}$\\(Hermite)} \\
	$L$ & $-\hat{E}_1$ &$\hat{D}_1+(a+1)\hat{E}_0$ & $-|\lambda|$ & \makecell{$P^{(L,a)}_{\lambda}$\\(Laguerre)}  \\
	$J$ & $-\hat{D}_2-(a+b+2)\hat{E}_1$& $\hat{D}_1+(a+1)\hat{E}_0$ & $-\sum_{(i,j)\in\lambda}2j+a+b+2\beta(N-i)$ &  \makecell{$P^{(J,a,b)}_{\lambda}$\\(Jacobi)}\\
	$B$ & $\hat{D}_2+(a+2)\hat{E}_1$ & $2\hat{E}_0$ & $\sum_{(i,j)\in\lambda}2j+a+2\beta(N-i)$ & \makecell{$P^{(B,a)}_{\lambda}$\\(Bessel)}\\
	\bottomrule[2pt]
\end{tabular}
\end{table*}

In order to further explore properties of the polynomials $P^{(X)}_{\lambda}$, we rewrite  (\ref{uptra}) as
\begin{eqnarray}\label{RelationXC}
	P^{(X)}_{\lambda}=c^{(X)}_{\lambda,\lambda} \mathcal{L}_{\lambda}^{(X)}P^{(C)}_{\lambda},
\end{eqnarray}
where the operators $\mathcal{L}^{(X)}_{\lambda}$ are given by
\begin{eqnarray}\label{LX}
	\mathcal{L}^{(X)}_{\lambda}=\prod_{\mu\subset\lambda}\frac{\mathcal{H}^{(X)}-\epsilon^{(X)}_{\mu}}{\epsilon^{(X)}_{\lambda}-\epsilon^{(X)}_{\mu}}.
\end{eqnarray}

It is easy to see that the operators $\prod_{\mu\subset\lambda} (\mathcal{H}^{(X)}-\epsilon^{(X)}_{\mu})$ always annihilate the subspace $\operatorname{Span}\{P_{\mu}^{(X)}\} _{\mu\subset\lambda}$
which is equivalent to $\operatorname{Span}\{P_{\mu}^{(C)}\} _{\mu\subset\lambda}$. Thus we have
\begin{eqnarray}
\mathcal{H}^{(X)}\mathcal{L}^{(X)}_{\lambda}P^{(C)}_{\lambda}=\mathcal{L}^{(X)}_{\lambda}\left(\mathcal{H}_0^{(X)}+\mathcal{H}_{<0}^{(X)}\right)P^{(C)}_{\lambda}
=\epsilon^{(X)}_{\lambda}\mathcal{L}^{(X)}_{\lambda}P^{(C)}_{\lambda}.
\end{eqnarray}

In particular, when $X\in\{H,L\}$, as the solutions of  $\mathcal{H}^{(X)}\mathcal{L}^{(X)}_{\lambda} =\mathcal{L}^{(X)}_{\lambda}\mathcal{H}_0^{(X)}$, $\mathcal{L}^{(X)}_{\lambda}$
can be expressed as  \cite{BakerCMP,Lassalle1,Lassalle2,Lassalle3,Sogo,Desrosiers}
\begin{subequations}\label{LHLL}
\begin{eqnarray}
	&&\mathcal{L}_{\lambda}^{(H)}=e^{-\mathcal{H}^{(H)}_{<0}/4}=e^{-\hat{D}_0/4}, \\
	&&\mathcal{L}_{\lambda}^{(L)}=e^{-\mathcal{H}^{(L)}_{<0}}=e^{-\hat{D}_1-(a+1)\hat{E}_0},
\end{eqnarray}
\end{subequations}
which no longer depend on the partition $\lambda$. Thus we denote $\mathcal{L}_{\lambda}^{(H)}=\mathcal{L}^{(H)}$ and $\mathcal{L}_{\lambda}^{(L)}=\mathcal{L}^{(L)}$.

Let us consider a generic differential operator $\mathcal{\tilde{H}}^{(C)}$ which comes from the hierarchy of commuting trigonometric CS operators and satisfies
$\mathcal{\tilde{H}}^{(C)}P^{(C)}_\lambda=\tilde{\epsilon}_\lambda P^{(C)}_\lambda$.
The operator $\mathcal{\tilde{H}}^{(C)}$ is not necessarily restricted to the original CS operator $\mathcal{H}^{(C)}$.
By means of $\mathcal{\tilde{H}}^{(C)}$ and $\mathcal{L}^{(X)}$,
we may construct the generalized CS operators $\mathcal{\tilde{H}}^{(X)}_{\lambda}$ for $X\in\{H,L,J,B\}$
\begin{eqnarray}\label{LNC}
\mathcal{\tilde{H}}^{(X)}_{\lambda}=\mathcal{L}^{(X)}_{\lambda}
\circ\mathcal{\tilde{H}}^{(C)}\circ\bar{\mathcal{L}}^{(X)}_{\lambda},
\end{eqnarray}
where the operators $\bar{\mathcal{L}}^{(X)}_{\lambda}$ satisfy $\bar{\mathcal{L}}^{(X)}_{\lambda}P^{(X)}_{\lambda}=c^{(X)}_{\lambda,\lambda}P^{(C)}_{\lambda}$.
Thus for the operators $\mathcal{\tilde{H}}^{(X)}_{\lambda}$, we have $\mathcal{\tilde{H}}^{(X)}_{\lambda}P^{(X)}_\lambda=\tilde{\epsilon}_\lambda P^{(X)}_\lambda.$

Especially, when $X \in \{H,L\}$ in (\ref{LNC}), the generalized CS operators can be written as
\begin{subequations}
\begin{eqnarray}
	\mathcal{\tilde{H}}^{(H)}&=&e^{-\hat{D}_0/4}\circ\mathcal{\tilde{H}}^{(C)}\circ e^{\hat{D}_0/4},\label{LNH}\\
	\mathcal{\tilde{H}}^{(L)}&=&e^{-\hat{D}_1-(a+1)\hat{E}_0}\circ\mathcal{\tilde{H}}^{(C)}\circ e^{\hat{D}_1+(a+1)\hat{E}_0},
\end{eqnarray}
\end{subequations}
which do not depend on the partition $\lambda$. Here $\bar{\mathcal{L}}^{(H)}_{\lambda} =\left(\mathcal{L}^{(H)}\right)^{-1}=e^{\hat{D}_0/4}$ and $\bar{\mathcal{L}}^{(L)}_{\lambda}=\left(\mathcal{L}^{(L)}\right)^{-1}=e^{\hat{D}_1+(a+1)\hat{E}_0}$.

It is noted that when $\mathcal{\tilde{H}}^{(C)}=-2\hat{E}_1$ in (\ref{LNH}), it gives the Lassalle-Nekrasov correspondence, which describes the relation between circular
and Hermite systems \cite{Lassalle1,Lassalle2,Lassalle3,Nekrasov}. There are more relations beyond the original Lassalle-Nekrasov correspondence in (\ref{LNC}). Hence we call (\ref{LNC})
the generalized Lassalle-Nekrasov correspondence.

\section{$\hat{W}$-operators and $\hat{O}$-operator}\label{WOoperator}
Let us recall the spherical degenerate double affine Hecke algebra ($\mathbf{SH}_N$). It is initially defined from degenerate double affine Hecke algebra of $GL_N$ ($\mathbf{H}_N$)\cite{Cherednik}
by the symmetrization $\mathbf{SH}_N=\mathbf{S}\circ\mathbf{H}_N\circ\mathbf{S}$, where $\mathbf{S}=\frac{1}{N!}\sum_{\sigma\in\mathfrak{S}_N} \sigma$ is the complete idempotent in $\mathbb{C}[\mathfrak{S}_N]$.

The $\mathbf{SH}_N$ algebra is defined by generators $\{D^{(N)}_{r,s}\}_{r\in\mathbb{Z},s\in\mathbb{Z}_{\ge0}}$ with commutation relations \cite{Schiffmann}
\begin{align}
	[D^{(N)}_{0,l},D^{(N)}_{\pm1,k}]&=\pm D^{(N)}_{\pm1,l+k-1}, & [D^{(N)}_{\pm 1,1},D^{(N)}_{\pm l,0}]
	&=\pm lD^{(N)}_{\pm(l+1),0},\nonumber \\  [D^{(N)}_{0,l+1},D^{(N)}_{\pm k,0}]&=\pm D^{(N)}_{\pm k,l},&
	[D^{(N)}_{0,l},D^{(N)}_{0,k}]&=0,\nonumber \\
	[D^{(N)}_{-1,l},D^{(N)}_{1,k}]&=\mathbf{E}^{(N)}_{l+k},& &
\end{align}
where $D^{(N)}_{0,0}=N$ and $\mathbf{E}^{(N)}_k$ are determined through the formula
\begin{eqnarray}
	1+(1-\beta)\sum_{l=0}^{\infty}\mathbf{E}^{(N)}_ls^{l+1}=\frac{(1+(1-\beta)s)(1+N\beta s)} {1+(1-\beta)s+N\beta s}\exp\left\{ \sum_{l=0}^{\infty}D^{(N)}_{0,l+1}\omega_{l}(s)\right\},
\end{eqnarray}
with
\begin{align}
\omega_{l}(s)&=\sum_{q=1,-\beta,\beta-1}s^l(\phi_l(1-qs)-\phi_l(1+qs)),\nonumber\\
\phi_{0}(s)&=-\log(s),\quad \phi_l(s)=(s^{-l}-1)/l\quad l\geq1.
\end{align}

The $N$-body representation of the $\mathbf{SH}_N$ algebra is given by \cite{Schiffmann,Cherednik}
\begin{eqnarray}
	D^{(N)}_{n,0}&=&p_n(\mathbf{x}),\qquad \forall n\in\mathbb{Z}, \nonumber\\
	D^{(N)}_{0,1}&=&\hat{E}_1,\nonumber\\
	D^{(N)}_{0,2}&=&{1\over2}\hat{D}_2+(1-\xi)\hat{E}_1,
\end{eqnarray}
where $p_n(\mathbf{x})=\sum_{j=1}^{N}x_j^n$, $\xi=(N-1)\beta+1$ in this paper. Other elements in $\mathbf{SH}_N$ algebra can be generated by the triples $(D^{(N)}_{0,2},D^{(N)}_{1,0},D^{(N)}_{-1,0})$.

Moreover, the Cartan generators $D^{(N)}_{0,l}$ for $l\ge1$ are defined by the linearization \cite{Schiffmann}
\begin{eqnarray}
D^{(N)}_{0,l}=\mathbf{S}\circ B^{(N)}_{l-1}(\pi_1-N\beta ,\cdots,\pi_N-N\beta)\circ\mathbf{S},
\end{eqnarray}
where $\pi_j=x_j\frac{\partial}{\partial x_j}+\beta\sum_{k\neq j}\frac{1}{1-x_k/x_j}(1-s_{j,k}) +\beta\sum_{k<j}s_{j,k}$ for $j=1,\dots,N$ are the Cherednik operators,
$s_{j,k}$ are the permutation operators of coordinates $x_j$ and $x_k$, and $B^{(N)}_{l}$ satisfy
$B^{(N)}_l(\lambda_1-\beta,\cdots,\lambda_N-N\beta) =\sum_{(i,j)\in\lambda}[j-1+(1-i)\beta]^{l-1}$.
Hence the actions of $D^{(N)}_{0,l}$ on Jack polynomials are
\begin{eqnarray}
D^{(N)}_{0,l}P^{(C)}_{\lambda}=\sum_{(i,j)\in\lambda}[j-1+(1-i)\beta]^{l-1}P^{(C)}_{\lambda}.
\end{eqnarray}

The circular CS operator $\mathcal{H}^{(C)}(\mathbf{x})$ is invariant under the coordinate transformation
$\mathbf{x}\mapsto\mathbf{x}^{-1}$.
Thus we may divide it into the following two parts
\begin{eqnarray}
	\mathcal{H}^{(C)}(\mathbf{x})=\hat{W}_0(u;\mathbf{x})+\hat{W}_0(u;\mathbf{x}^{-1}),
\end{eqnarray}
which holds for any parameter $u\in\mathbb{C}$ and
\begin{eqnarray}
	\hat{W}_0(u;\mathbf{x})=D^{(N)}_{0,2}+uD^{(N)}_{0,1}=\frac{1}{2}\hat{D}_2+(u-\xi+1)\hat{E}_1.
\end{eqnarray}

In terms of $\hat{W}_0$, we construct the operators
\begin{subequations}\label{W-W+}
\begin{eqnarray}
	&&\hat{W}_{1}^{(1)}(u;\mathbf{x})=\mathbf{ad}_{\hat{W}_0(u;\mathbf{x})}\left(p_1(\mathbf{x})\right)=\hat{E}_2+up_1(\mathbf{x}),\\
	&&\hat{W}_{-1}^{(1)}(u;\mathbf{x})=-\mathbf{ad}_{\hat{W}_0(u;\mathbf{x})}\left(p_{-1}(\mathbf{x})\right)=\hat{E}_0+(u-\xi)p_{-1}(\mathbf{x}),
\end{eqnarray}
\end{subequations}
where $\mathbf{ad}_f\ g:=[f,g]$.

Then for any $v\in\mathbb{C}$, we have
\begin{eqnarray}
	\sum_{n=1}^{\infty}p_{\pm n}(\mathbf{x})t^{n-1}=\exp\left(\pm t\mathbf{ad}_ {\hat{W}_{\pm 1}^{(1)}(v;\mathbf{x})}\right) \left(p_{\pm 1}(\mathbf{x})\right).
\end{eqnarray}

Furthermore, we construct two families of operators $\hat{W}^{(p)}_{n}$ and $\hat{W}^{(p)}_{-n}$ by
\begin{eqnarray}\label{sumWt}
\sum_{n=1}^{\infty}\hat{W}_{\pm n}^{(p)}(\mathbf{a};\mathbf{x})t^{n-1}=
\exp\left(\pm t\mathbf{ad}_{\hat{W}_{\pm 1}^{(p+1)} (v,\mathbf{a};\mathbf{x})}\right)\hat{W}_{\pm 1}^{(p)}(\mathbf{a};\mathbf{x}),
\end{eqnarray}
where $p\in\mathbb{N}$, $\mathbf{a}=(a_1,a_2,\cdots,a_p)$ and
\begin{eqnarray}\label{W1pr}
&&\hat{W}_{\pm 1}^{(p)}(\mathbf{a};\mathbf{x})=\prod_{k=1}^{p}\mathbf{ad}_{\pm \hat{W}_0(a_k;\mathbf{x})}p_{\pm 1}(\mathbf{x}).
\end{eqnarray}

We see that the operators (\ref{CSoperator}) and (\ref{LHLL}) can be represented by the $\hat W$-operators
\begin{subequations}
\begin{eqnarray}
\mathcal{H}^{(R)}(\mathbf{x})&=& \hat{W}^{(1)}_{-2}(\xi;\mathbf{x})\\ \mathcal{H}^{(C)}(\mathbf{x})&=&\hat{W}_{0}(u;\mathbf{x})
+\hat{W}_{0}(u;\mathbf{x}^{-1}),\qquad \forall u\in\mathbb{C},\\
\mathcal{H}^{(H)}(\mathbf{x})&=& \hat{W}^{(1)}_{-2}(\xi;\mathbf{x})-2\hat{E}_1
=-2e^{-\frac{1}{4}\hat{W}^{(1)}_{-2}(\xi;\mathbf{x})}\hat{E}_1
e^{\frac{1}{4}\hat{W}^{(1)}_{-2}(\xi;\mathbf{x})},\\
	\mathcal{H}^{(L)}(\mathbf{x})&=& \hat{W}^{(2)}_{-1} (\xi,a+\xi;\mathbf{x})-\hat{E}_1
=-e^{-\hat{W}^{(2)}_{-1}(\xi,a+\xi;\mathbf{x})}\hat{E}_1
e^{\hat{W}^{(2)}_{-1}(\xi,a+\xi;\mathbf{x})},\\
	\mathcal{H}^{(J)}(\mathbf{x})&=&
	\hat{W}^{(2)}_{-1}(\xi,a+\xi;\mathbf{x})-\left(\hat{W}_{0}
(a+\xi;\mathbf{x})+\hat{W}_{0}(b+\xi;\mathbf{x})\right),\\
	\mathcal{H}^{(B)}(\mathbf{x})&=&
	\hat{W}_{0}(a+\xi;\mathbf{x})+\hat{W}_{0}(\xi;\mathbf{x})+2\hat{W}^{(1)}_{-1}
(\xi;\mathbf{x}),\\
\mathcal{L}^{(H)}&=&\exp\left(-\hat{W}_{-2}^{(1)}(\xi;2\mathbf{x})\right),
\label{LHLL2}\\ \mathcal{L}^{(L)}&=&\exp\left(-\hat{W}_{-1}^{(2)}(\xi,a+\xi;\mathbf{x})\right),
\end{eqnarray}
\end{subequations}
where $\hat{E}_1= \frac{1}{\xi}(\hat{W}_{0}(\xi;\mathbf{x}) -\hat{W}_{0}(\xi;\mathbf{x}^{-1}))$.

Let us now construct the $\hat{O}$-operator by the generators of $\mathbf{SH}_{N}$ algebra
\begin{eqnarray}\label{Odefine}
	\hat{O}(a;\mathbf{x})=\exp\left\{D^{(N)}_{0,1}\cdot\log a -\sum_{n=1}^{\infty}
	\frac{(-a^{-1})^n}{n} D^{(N)}_{0,n+1}\right\},
\end{eqnarray}
such that
\begin{eqnarray}\label{OPEP}
\hat{O}(a;\mathbf{x})P^{(C)}_{\lambda}(\mathbf{x})= [a]_{\lambda}^{(\beta)}P^{(C)}_{\lambda}(\mathbf{x}),
\end{eqnarray}
where $[a]_{\lambda}^{(\beta)}=\prod_{(i,j)\in\lambda} [a+(j-1)+\beta(1-i)]$.

Let us consider the action of $\hat{W}_{\pm 1}^{(p)}(\mathbf{a};\mathbf{x})$ (\ref{W1pr}) on Jack polynomials
\begin{eqnarray}
	\hat{W}_{\pm 1}^{(p)}(\mathbf{a};\mathbf{x}) P_{\lambda}^{(C)}(\mathbf{x})&=&\sum_{\mu=\lambda\pm\square}
	R^{\pm}_{\lambda,\mu}\prod_{k=1}^{p}[a_k+j_{\square}-1+(1-i_{\square})\beta]P_{\mu}^{(C)}(\mathbf{x})\nonumber\\
	&=&\left(\hat{O}(\mathbf{a};\mathbf{x})\right)^{\pm 1}\circ p_{\pm 1}(\mathbf{x})\circ\left(\hat{O}(\mathbf{a;x}) \right)^{\mp 1}P_{\lambda}^{(C)}(\mathbf{x}),
\end{eqnarray}
where $\hat{O}(\mathbf{a;x})=\prod_{k=1}^p\hat{O} (a_k;\mathbf{x})$ and the coefficients $R^{\pm}_{\lambda,\mu}$ are given by the Pieri formulas for Jack polynomials
$p_{\pm1}(\mathbf{x}) P_{\lambda}^{(C)}(\mathbf{x}) =\sum_{\mu=\lambda\pm\square} R^{\pm}_{\lambda,\mu}P_{\mu}^{(C)}(\mathbf{x})$.
It implies that
\begin{eqnarray}
	\hat{W}_{\pm 1}^{(p)}(\mathbf{a};\mathbf{x}) =\left(\hat{O}(\mathbf{a;x})\right)^{\pm 1}\circ p_{\pm 1}(\mathbf{x})\circ\left(\hat{O}(\mathbf{a;x}) \right)^{\mp 1}.
\end{eqnarray}
Then it follows from (\ref{sumWt}) that the $\hat{W}$-operators $\hat{W}^{(p)}_{\pm n}$ can be expressed as
\begin{eqnarray}\label{Wops}
	\hat{W}_{\pm n}^{(p)}(\mathbf{a;x}) =\left(\hat{O}(\mathbf{a};\mathbf{x})\right)^{\pm 1}\circ p_{\pm n}(\mathbf{x}) \circ \left(\hat{O}(\mathbf{a;x})\right)^{\mp 1}.
\end{eqnarray}
It is worth noting that the constructions of $\hat W$-operators (\ref{Wops}) are based on the abelian algebraic relation $[p_{n}(\mathbf{x}),p_{m}(\mathbf{x})]=0$ for any $n,m\in\mathbb{Z}$,
but not the Heisenberg algebra. Thus the $\hat W$-operators (\ref{Wops}) are different from those in Refs.\cite{Rui22,Fan}.

From (\ref{Wops}), it is easy to obtain the commutation relations
\begin{eqnarray}\label{Commu}
[\hat{W}_{n}^{(p)}(\mathbf{a;x}),\hat{W}_{m}^{(p)}(\mathbf{a;x})]=0,\qquad [\hat{W}_{-n}^{(p)}(\mathbf{a;x}),\hat{W}_{-m}^{(p)}(\mathbf{a;x})]=0,
\end{eqnarray}
where $n,m\in\mathbb{Z}_+$ and $p\in\mathbb{N}$.

Furthermore, the infinite sets of commutative $\hat{W}$-operators $\hat{W}_{n}^{(p)}(\mathbf{a;x})$ and $\hat{W}_{-n}^{(p)}(\mathbf{a;x})$ may be regarded as the Hamiltonians of certain many-body
integrable systems beyond the ordinary CS-type models \cite{MirCom}.

For example, the operators $\hat{W}^{(1)}_{-n}$ are the higher rational CS operators \cite{MirMany}
\begin{eqnarray} \hat{W}^{(1)}_{-n}(\xi;\mathbf{x})=\mathcal{H}_n^{(R)}=\sum_{r=1}^{n}\binom{n}{r}\beta^{n-r}\sum_{j=1}^{N}
\left(\sum_{\overset{K\subseteq\{1,\cdots,N\}\setminus j}{|K|=n-r}}\prod_{k\in K}\frac{1}{x_j-x_k}\right) \frac{\partial^r}{\partial x_j^r},\quad n\le N,
\end{eqnarray}
and the operators $\hat{W}^{(1)}_{n}(\xi;\mathbf{x})$ can be given by \cite{MirMany}
\begin{eqnarray}
\hat{W}^{(1)}_{n}(\xi;\mathbf{x})=-(\prod_{j=1}^{N}x_j)^{-\xi}\circ \hat{W}^{(1)}_{-n}(\xi; \mathbf{x}^{-1}) \circ(\prod_{j=1}^{N}x_j)^{\xi}.
\end{eqnarray}

Let us introduce the following operators
\begin{eqnarray}\label{QpmC} \hat{Q}^C_{\pm}(\mathbf{x})=\hat{W}_{\pm1}^{(p)}(\mathbf{0;x})+\sum_{k=1}^{p}C_k\hat{W}_{\pm1}^{(p-k)}(\mathbf{0;x}),
\end{eqnarray}
where the coefficients $C_k\in\mathbb{R}$. The operators $\hat{Q}^C_{\pm}(\mathbf{x})$ belong to the cone type in Ref.\cite{MirCom}.

Based on the decomposition of the $\hat W$-operators
\begin{eqnarray}\label{compo}
\hat{W}_{\pm1}^{(p)}(\mathbf{a;x})=\hat{W}_{\pm1} ^{(p)}(\mathbf{0;x})+\sum_{k=1}^{p}e_k(\mathbf{a})\hat{W}_{\pm1}^{(p-k)}(\mathbf{0;x}),
\end{eqnarray}
where $e_k(\mathbf{a})=\sum_{1\le i_1<\cdots<i_k\le p}a_{i_1}a_{i_2}\cdots a_{i_k}$ are elementary symmetric polynomials, the operators (\ref{QpmC}) can always be rewritten as
\begin{eqnarray}\label{Qpmc1}
	\hat{Q}^C_{\pm}(\mathbf{x})= \hat{W}_{\pm1}^{(p)}(\mathbf{a_c;x}),
\end{eqnarray}
where the parameters $\mathbf{a_c}\in\mathbb{C}^p$ satisfy $(e_1(\mathbf{a_c}),\cdots,e_p(\mathbf{a_c})) =(C_1,\cdots,C_p)$.

From (\ref{Qpmc1}), we see that the operators $\hat{Q}^C_{\pm}(\mathbf{x})$ constructed by the linear combination (\ref{QpmC}) also belong to the hierarchy of $\hat{W}$-operators. By (\ref{Commu}),
the operators $\hat{Q}^C_{\pm}(\mathbf{x})$ can be regarded as the Hamiltonians of certain many-body integrable systems.

\section{Generalized hypergeometric functions}\label{Hyper}
\subsection{$O$-representations of generalized hypergeometric functions}
The hypergeometric function with single variable is given by \cite{Erdelyi}
\begin{eqnarray}
	_pF_r(\mathbf{a;b};x)=\sum_{n=0}^{\infty}\frac{\prod_{j=1}^{p} (a_j)_n}{\prod_{k=1}^{r}(b_k)_n} \frac{x^n}{n!},
\end{eqnarray}
where $\mathbf{a}=(a_1,\cdots,a_p)\in\mathbb{C}^p$, $\mathbf{b}=(b_1,\cdots,b_r)\in\mathbb{C}^r$ and $(a)_n=\prod_{s=0}^{n-1}(a+s)$.
It can also be defined as the solution of the differential equation
\begin{eqnarray}\label{Consf} \left(\frac{d}{dx}\prod_{k=1}^{r}(x\frac{d}{dx}+b_k-1)
-\prod_{j=1}^{p}(x\frac{d}{dx}+a_j)\right){_pF_r}(\mathbf{a;b};x)=0.
\end{eqnarray}

The hypergeometric functions with one and two sets of variables are defined as \cite{Koranyi,Macdonald13,Kaneko}
\begin{subequations}\label{defHyper}
\begin{eqnarray}
_p\mathcal{F}^{(\beta)}_r(\mathbf{a;b};\mathbf{x;z})&=&\sum_{\lambda}\frac{1}{|\lambda|!} \frac{\prod_{j=1}^{p}[a_j]^{(\beta)}_{\lambda}}{\prod_{k=1}^{r}[b_k]^{(\beta)}_{\lambda}}
\frac{P^{(C)}_{\lambda}(\mathbf{x})P^{(C)}_{\lambda}(\mathbf{z})}{P^{(C)}_{\lambda}
(1^N)},\label{defineFpr}\\
_pF^{(\beta)}_r(\mathbf{a;b};\mathbf{x})&=&_p\mathcal{F}^{(\beta)}_r(\mathbf{a;b};
\mathbf{x};1^N),\label{definefpr}
\end{eqnarray}
\end{subequations}
where $\mathbf{x}=(x_1,x_2,\cdots,x_N)$, $\mathbf{z}=(z_1,z_2,\cdots,z_N)$ and $\mathbf{z}=1^N$ means $z_i=1$ for $i=1,\cdots,N$.
We call (\ref{defHyper}) generalized hypergeometric functions in this paper.

We write down a few special cases of (\ref{defHyper})
\begin{subequations}
\begin{align}
	_0F^{(\beta)}_0(\mathbf{x})&=e^{p_1(\mathbf{x})},\\
	_0\mathcal{F}^{(\beta)}_0(\mathbf{x};\mathbf{z})&= \sum_{\lambda}\frac{1}{|\lambda|!} \frac{P^{(C)}_{\lambda}(\mathbf{x})P^{(C)}_{\lambda}(\mathbf{z})}
	{P^{(C)}_{\lambda}(1^N)}=\sum_{\lambda}\frac{1}{[N\beta]_{\lambda}^{(\beta)}} \frac{J_{\lambda}(\mathbf{x})J_{\lambda}(\mathbf{z})}{j_{\lambda}},\\
	{_1\mathcal{F}^{(\beta)}_0}(N\beta;\mathbf{x;z})&=	\sum_{\lambda}\frac{J_{\lambda}(\mathbf{x})J_{\lambda}(\mathbf{z})}
	{j_{\lambda}}=\exp\left(\beta\sum_{n=1}^{\infty}\frac{p_n(\mathbf{x})p_n(\mathbf{z})}{n}\right),\label{F10}
\end{align}
\end{subequations}
where (\ref{F10}) is the $\beta$-deformed Cauchy identity \cite{Macdonald,Stanley89} and we have used
$J_{\lambda}(1^N)= \beta^{-|\lambda|}[N\beta]_{\lambda}^{(\beta)}$.

With the help of the $\hat{O}$-operator (\ref{Odefine}), we may express the generalized hypergeometric functions (\ref{defHyper}) as
\begin{subequations}\label{OrepF}
\begin{eqnarray}
_p\mathcal{F}^{(\beta)}_r(\mathbf{a;b};\mathbf{x};\mathbf{z})&=&\hat{O}^{-1}(\mathbf{b;x})
\hat{O}(\mathbf{a;x}){_0\mathcal{F}^{(\beta)}_0(\mathbf{x;z})},\\
_pF^{(\beta)}_r(\mathbf{a;b};\mathbf{x})&=& \hat{O}^{-1}(\mathbf{b;x})\hat{O}(\mathbf{a;x}){_0F^{(\beta)}_0(\mathbf{x})}.
\end{eqnarray}
\end{subequations}
We call (\ref{OrepF}) the $O$-representations of generalized hypergeometric functions.

Moreover, for special cases of the generalized hypergeometric functions, we can derive their $W$-representations from (\ref{Wops}) and (\ref{OrepF})
\begin{subequations}
\begin{eqnarray}
	_{p+1}\mathcal{F}^{(\beta)}_0(N\beta,\mathbf{a};\mathbf{x;z})&=&\hat{O}(\mathbf{a};\mathbf{x})\exp\left(\beta\sum_{n=1}^{\infty}
	\frac{p_n(\mathbf{x})p_n(\mathbf{z})}{n}\right)\label{Wrep1}\nonumber\\ &=&\exp\left(\beta\sum_{n=1}^{\infty}\frac{\hat{W}_{n}^{(p)}(\mathbf{a;x})p_n(\mathbf{z})}{n}\right)\cdot1,\\
	_{p}F^{(\beta)}_0(\mathbf{a};\mathbf{x})&=&\hat{O}(\mathbf{a};\mathbf{x})
	e^{p_1(\mathbf{x})}=e^{\hat{W}_{1}^{(p)}(\mathbf{a;x})}\cdot1.\label{Wrep2}
\end{eqnarray}
\end{subequations}

\subsection{Constraints of the generalized hypergeometric function $_p\mathcal{F}_r(\mathbf{a;b;x;z})$}

Let us consider the generalized hypergeometric functions with two sets of variables. It is easy to check that
\begin{eqnarray}\label{RW-W+} &&\hat{W}^{(1)}_{1}(N\beta;\mathbf{z}){_1\mathcal{F}^{(\beta)}_0}(N\beta;\mathbf{x;z})\nonumber\\
&=&\left[\beta\sum_{n>0}p_{n+1}(\mathbf{z})p_n(\mathbf{x})+N\beta p_1(\mathbf{z})\right]\exp\left( \beta\sum_{n=1}^{\infty}\frac{p_n(\mathbf{x})p_n(\mathbf{z})}{n}\right)\nonumber\\
&=&\hat{W}^{(1)}_{-1}(\xi;\mathbf{x}){_1\mathcal{F}^{(\beta)}_0}(N\beta;\mathbf{x;z}),
\end{eqnarray}
where we have used the collective coordinate representations
\begin{align}
	\hat{W}^{(1)}_{1}(N\beta;\mathbf{z})&=\hat{E}_2(\mathbf{z})=\sum_{n>0}np_{n+1}(\mathbf{z})\frac{\partial}{\partial p_n(\mathbf{z})}+N\beta p_1(\mathbf{z}),\nonumber\\
	\hat{W}^{(1)}_{-1}(\xi;\mathbf{x})&=\hat{E}_0(\mathbf{x})=\sum_{n>0}(n+1)p_n(\mathbf{x})\frac{\partial}{\partial p_{n+1}(\mathbf{x})}+N\frac{\partial}{\partial p_1(\mathbf{x})}.
\end{align}
The actions of $\hat{O}(\mathbf{a;z}) \hat{O}^{-1}(N\beta;\mathbf{z})\hat{O}^{-1}(\mathbf{b;x})$ and $\hat{O}(v,\mathbf{a;z})\hat{O}^{-1}(N\beta;\mathbf{z})\hat{O}^{-1}(v,\mathbf{b;x})$
on both sides of (\ref{RW-W+}) respectively give
\begin{eqnarray} \label{WpWr}
\hat{W}^{(p)}_{1}(\mathbf{a;z}){_p\mathcal{F}^{(\beta)}_r}(\mathbf{a;b;x;z})&=&\hat{W}^{(r+1)}_{-1}
(\xi;\mathbf{b;x}){_p\mathcal{F}^{(\beta)}_r}(\mathbf{a;b;x;z}),\nonumber\\ \hat{W}^{(p+1)}_{1}(v,\mathbf{a;z}){_p\mathcal{F}^{(\beta)}_r}(\mathbf{a;b;x;z})&=&\hat{W}^{(r+2)}_{-1}(v,\xi;
\mathbf{b;x}){_p\mathcal{F}^{(\beta)}_r}(\mathbf{a;b;x;z}).
\end{eqnarray}

By means of (\ref{sumWt}) and (\ref{WpWr}), we have
\begin{eqnarray}
	&&\hat{W}^{(p)}_{n}(\mathbf{a;z}){_p\mathcal{F}^{(\beta)}_r}(\mathbf{a;b;x;z})\nonumber\\
	&=&\frac{1}{(n-1)!}\left(\mathbf {ad}^{n-1}_{\hat{W}^{(p+1)}_{1}(v,\mathbf{a;z})}\right) \hat{W}^{(p)}_{1}(\mathbf{a;z}) {_p\mathcal{F} ^{(\beta)}_r}(\mathbf{a;b;x;z})\nonumber\\
	&=&\frac{(-1)^{n-1}}{(n-1)!}\left(\mathbf {ad}^{n-1}_{\hat{W}^{(r+2)}_{-1}(v,\xi,\mathbf{b;x})}\right) \hat{W}^{(r+1)}_{-1}(\xi,\mathbf{b;x}){_p\mathcal{F} ^{(\beta)}_r}(\mathbf{a;b;x;z})\nonumber\\
	&=&\hat{W}^{(r+1)}_{-n}(\xi,\mathbf{b;x}){_p\mathcal{F}^{(\beta)}_r}(\mathbf{a;b;x;z}).
\end{eqnarray}
Thus we obtain the constraints
for the generalized hypergeometric function ${_p\mathcal{F}^{(\beta)}_r}(\mathbf{a;b;x;z})$
\begin{eqnarray}\label{eigenWF} \left(\hat{W}^{(p)}_{n}(\mathbf{a;z})-\hat{W}^{(r+1)}_{-n}
(\xi,\mathbf{b;x})\right){_p\mathcal{F}^{(\beta)}_r}(\mathbf{a;b;x;z})=0.
\end{eqnarray}

When specialized to the case of $(p,r,n)=(0,0,2)$ in (\ref{eigenWF}), we have
\begin{eqnarray}\label{F00cons}
	\left(p_2(\mathbf{z})-W_{-2}^{(1)}(\xi;\mathbf{x})\right){_0\mathcal{F}}^{(\beta)}_0(\mathbf{x};\mathbf{z})=0.
\end{eqnarray}

We notice that the $\beta$-deformed HCIZ integral also satisfies \cite{Eynard}
\begin{eqnarray}
	\left(p_2(\mathbf{z})-W_{-2}^{(1)}(\xi;\mathbf{x})\right)I_{IZ}(\mathbf{x,z})=0,
\end{eqnarray}
where
\begin{eqnarray}
	I_{IZ}(\mathbf{x,z})=I_{IZ}(X,Z)
	=\int [dU]_{\beta} e^{TrXUZU^{\dagger}},
\end{eqnarray}
$X=\mathbf{diag}(x_1,\cdots,x_N)$, $Z=\mathbf{diag}(z_1,\cdots,z_N)$ and $[dU]_{\beta}$ is the Haar measure on certain compact Lie group related to the parameter $\beta$.
Thus we recover the relation \cite{MirBGW}
\begin{eqnarray}\label{F00IZ}
	_0\mathcal{F}^{(\beta)}_0(\mathbf{x};\mathbf{z})\sim I_{IZ}(\mathbf{x,z}),
\end{eqnarray}
where $``\sim"$ means that the left-hand side is equivalent to the right-hand side up to a constant factor in this paper. The relation (\ref{F00IZ}) will be used to construct certain integral chains in the next section. Furthermore, by (\ref{eigenWF}) and (\ref{F00IZ}), we give the general constraints for the $\beta$-deformed HCIZ integral
\begin{eqnarray}
\left(p_n(\mathbf{z})-W_{-n}^{(1)}(\xi;\mathbf{x})\right)I_{IZ}(\mathbf{x,z})=0.
\end{eqnarray}

Let us discuss the properties of the generalized hypergeometric functions $_p\mathcal{F}_r(\mathbf{a;b;x;z})$ and Hermite or Laguerre polynomials and provide the proofs
for them based on the previous statements.

(I) Taking the action of $\hat{O}(\xi;\mathbf{x})$ on (\ref{eigenWF}) with $(p,r)=(1,0)$, we obtain that
\begin{eqnarray}\label{fzfx}
	f(\mathbf{z}){_1\mathcal{F}^{(\beta)}_0} (\xi;\mathbf{x;z})=f(\mathbf{x}^{-1}){_1\mathcal{F}^{(\beta)}_0}(\xi;\mathbf{x;z}),
\end{eqnarray}
for any symmetric polynomial $f$, where we make use of the fact that $p_{\lambda}=p_{\lambda_1} p_{\lambda_2}\cdots$ form a basis of symmetric polynomial space.

(II) By (\ref{RelationXC}) and (\ref{LHLL2}),
we immediately obtain the Hermite and Laguerre polynomials \cite{Lassalle1,Lassalle2,Lassalle3}
\begin{subequations}\label{PHPLPC}
\begin{eqnarray}
	P_{\lambda}^{(H)}(\mathbf{x})&=&e^{-\frac{1}{2}\hat{W}_{-2}^{(1)}(\xi;\mathbf{x})}\frac{P_{\lambda}^{(C)}(\mathbf{x})}{P_{\lambda}^{(C)}(1^N)},\\
	 P^{(L,a)}_{\lambda}(\mathbf{x})&=&e^{-\hat{W}_{-1}^{(2)}(\xi,a+\xi;\mathbf{x})}\frac{P_{\lambda}^{(C)}(-\mathbf{x})}{|\lambda|!P_{\lambda}^{(C)}(1^N)} \label{PLPC}.
\end{eqnarray}
\end{subequations}

(III) Taking $(p,r,n)=(0,0,2)$ and $(p,r,n)=(0,1,1)$ in (\ref{eigenWF}), respectively, it is not difficult to obtain that \cite{BakerCMP}
\begin{subequations}\label{p2p1F}
\begin{eqnarray}
	e^{-p_2(\mathbf{z})}{_0\mathcal{F}^{(\beta)}_0} (2\mathbf{x};\mathbf{z})&=&\sum_{\lambda}\frac{1}{|\lambda|!}P^{(H)}_{\lambda}(\mathbf{x})P^{(C)}_{\lambda}(\mathbf{z}),\label{p2F00}\\
	e^{p_1(\mathbf{z})}{_0\mathcal{F}^{(\beta)}_1} (a+\xi;-\mathbf{x};\mathbf{z})&=&\sum_{\lambda}\frac{1}{[a+\xi]_{\lambda}^{(\beta)}}P^{(L,a)}_{\lambda}(\mathbf{x})P^{(C)}_{\lambda}(\mathbf{z})\label{p1F01}.
\end{eqnarray}
\end{subequations}

In particular, by taking $x_i=0$ for $i=1,\cdots,N$ in (\ref{p2p1F}), we can obtain the values of Hermite and Laguerre polynomials at zero
\begin{subequations}\label{zero}
\begin{eqnarray}
	P_{\lambda}^{(H)}(0^N)&=& (-1)^{|\lambda|/2}\beta^{|\lambda|} J_{\lambda}\{\beta^{-1}\delta_{n,2}\},\\
	P_{\lambda}^{(L,a)}(0^N)&=&\beta^{|\lambda|}(|\lambda|!)^{-1} J_{\lambda}\{\beta^{-1}\delta_{n,1}\} [a+\xi]^{(\beta)}_{\lambda},
\end{eqnarray}
\end{subequations}
where $J_{\lambda}\{\beta^{-1}\delta_{n,k}\} =J_{\lambda}\{p_n(\mathbf{x})=\beta^{-1}\delta_{n,k}\}$ are defined by the Cauchy identities
\begin{eqnarray}\label{Cauchy}
	e^{p_k(\mathbf{z})}=\sum_{m=0}^{\infty}\sum_{\lambda\vdash km}\frac{\beta^{|\lambda|} J_{\lambda}\{\beta^{-1}\delta_{n,k}\}}{|\lambda|!} P_{\lambda}^{(C)}(\mathbf{z}),\quad k\in\mathbb{N}.
\end{eqnarray}

(IV) Taking the action of $\hat{O}(c+\xi;\mathbf{z})$ on (\ref{PLPC}) and using $\hat{W}_1^{(1)}(c+\xi;\mathbf{z}) =\hat{E}_2(\mathbf{z})+cp_1(\mathbf{z})$,
$\hat{E}_2(\mathbf{z})=-\hat{E}_0(\mathbf{z}^{-1})$ and $e^{\hat{E}_0(\mathbf{z})}f(\mathbf{z})=f(1+\mathbf{z})e^{\hat{E}_0(\mathbf{z})}$,
we recover the formula \cite{BakerCMP}
\begin{eqnarray}\label{expansion1}
	&&{_1\mathcal{F}^{(\beta)}_1}(c+\xi;a+\xi; -\mathbf{x};\frac{1}{\mathbf{z}^{-1}-1})\prod_{j=1}^{N}(1-z_j)^{-c-\xi}\nonumber\\
	&=&\sum_{\lambda}\frac{[c+\xi]_{\lambda}^{(\beta)}}{[a+\xi]_{\lambda}^{(\beta)}}P^{(L,a)}_{\lambda}(\mathbf{x})P^{(C)}_{\lambda}(\mathbf{z}).
\end{eqnarray}
Moreover, taking $c=a$, $y_j={1\over1-z_j}$ for $j=1,2,\cdots,N$ in (\ref{expansion1}) and using the constraints (\ref{eigenWF}) for $(p,r,n)=(0,0,1)$, we have
\begin{eqnarray}\label{expansion2}
	e^{p_1(\mathbf{x})} {_0\mathcal{F}^{(\beta)}_0} (\mathbf{-x;y})\prod_{j=1}^{N}y_j^{a+\xi}=\sum_{\lambda}P^{(L,a)}_{\lambda}(\mathbf{x})P^{(C)}_{\lambda}(1-\mathbf{y}^{-1}).
\end{eqnarray}

(V) Taking $(p,r,n)=(0,0,1)$ in (\ref{eigenWF}), we recover the formula in Ref.\cite{BakerCMP}
\begin{eqnarray}\label{p1F00}
	e^{tp_1(\mathbf{z})} {_0\mathcal{F}^{(\beta)}_0}(\mathbf{x;z}) &=&e^{t\hat{W}_{-1}^{(1)}(\xi;\mathbf{x})} {_0\mathcal{F}^{(\beta)}_0}(\mathbf{x;z})\nonumber\\
	&=& \sum_{\lambda}\frac{P^{(C)}_{\lambda} (\mathbf{z})}{|\lambda|!}\sum_{\mu\subseteq\lambda}t^{|\lambda-\mu|}\binom{\lambda}{\mu}
	\frac{P^{(C)}_{\mu}(\mathbf{x}) }{P^{(C)}_{\mu}(1^N)},
\end{eqnarray}
where we have used $\hat{W}_{-1}^{(1)}(\xi;\mathbf{x}) =\hat{E}_0$, and the generalized binomial coefficient $\binom{\lambda}{\mu}$ is defined by \cite{Lassalle1,Lassalle2,Lassalle3}
\begin{eqnarray}\label{definebino}
	\frac{P^{(C)}_{\lambda}(t+\mathbf{x})}{P^{(C)}_{\lambda}(1^N)} =\sum_{\mu\subseteq\lambda} t^{|\lambda-\mu|}\binom{\lambda}{\mu}
	\frac{P^{(C)}_{\mu}(\mathbf{x})}{P^{(C)}_{\mu}(1^N)}.
\end{eqnarray}
Then by comparing the coefficient of $P^{(C)}_{\lambda}(\mathbf{x})$ in both sides of (\ref{p1F00}), we obtain that
\begin{eqnarray}\label{pPC}
	 e^{tp_1(\mathbf{z})}P^{(C)}_{\mu}(\mathbf{z})=\sum_{\lambda\supseteq\mu}t^{|\lambda-\mu|}\binom{\lambda}{\mu}\frac{|\mu|!}{|\lambda|!}P^{(C)}_{\lambda}(\mathbf{z}).
\end{eqnarray}

(VI) The transition matrix $c_{\lambda,\mu}^{(L)}$ in (\ref{uptra}) and its inverse are given by \cite{Lassalle1,Lassalle2,Lassalle3}
\begin{eqnarray}
P^{(L,a)}_{\lambda}(\mathbf{x})&=&\sum_{\mu\subseteq\lambda}(-1)^{|\mu|}\frac{[a+\xi]_{\lambda}  ^{(\beta)}}{[a+\xi]_{\mu}^{(\beta)}}
\binom{\lambda}{\mu}\frac{P^{(C)}_{\mu}(\mathbf{x})}{|\lambda|!P^{(C)}_{\mu}(1^N)}, \label{PLtoPC}\\
P^{(C)}_{\lambda}(\mathbf{x})&=&P^{(C)}_{\lambda}(1^N)\sum_{\mu\subseteq\lambda}(-1)^{|\mu|} \frac{[a+\xi]_{\lambda}^{(\beta)}}{[a+\xi]_{\mu}^{(\beta)}}
\binom{\lambda}{\mu}|\mu|!P^{(L,a)}_{\mu} (\mathbf{x}).\label{PCtoPL}
\end{eqnarray}

Let us now give a new proof of (\ref{PLtoPC}) and (\ref{PCtoPL}).
It follows from (\ref{PLPC}) that
\begin{eqnarray}
	P^{(L,a)}_{\lambda}(\mathbf{x})
	&=&\hat{O}^{-1}(a+\xi;\mathbf{x})\circ e^{-\hat{W}_{-1}^{(1)}(\xi;\mathbf{x})}\circ\hat{O}(a+\xi;\mathbf{x})\frac{P^{(C)}_{\lambda}(-\mathbf{x})}{|\lambda|!P^{(C)}_{\lambda}(1^N)}\nonumber\\
	&=&\hat{O}^{-1}(a+\xi;\mathbf{x}) \frac{(-1)^{|\lambda|}[a+\xi]_{\lambda}^{(\beta)}}{|\lambda|!}\frac{P^{(C)}_{\lambda}(\mathbf{x}-1)}{P^{(C)}_{\lambda}(1^N)}\nonumber\\
	&=&\sum_{\mu\subseteq\lambda}(-1)^{|\mu|}\frac{[a+\xi]_{\lambda}^{(\beta)}}{[a+\xi]_{\mu}^{(\beta)}}\binom{\lambda}{\mu}\frac{P^{(C)}_{\mu}(\mathbf{x})}{|\lambda|!P^{(C)}_{\mu}(1^N)},
\end{eqnarray}
where (\ref{definebino}) has been used in the last line. Thus (\ref{PLtoPC}) holds.

On the other hand, taking $(p,r,n)=(0,1,1)$ in (\ref{eigenWF}), it gives
\begin{eqnarray}\label{epF}
_0\mathcal{F}_{1}^{(\beta)}(a+\xi;\mathbf{x;z})=e^{-\hat{W}_{-1}^{(2)}(a+\xi;\mathbf{x})}e^{p_1(\mathbf{z})}
{_0\mathcal{F}^{(\beta)}_1}(a+\xi;\mathbf{x;z}).
\end{eqnarray}
Expanding both sides of (\ref{epF}) and using (\ref{PLPC}), we obtain
\begin{eqnarray}\label{PCPLPC}
\sum_{\lambda}\frac{P^{(C)}_{\lambda}(\mathbf{x})P^{(C)}_{\lambda} (\mathbf{z})}{P^{(C)}_{\lambda}(1^N) |\lambda|![a+\xi]_{\lambda}^{(\beta)}}=\sum_{\mu\subseteq\lambda} (-1)^{|\mu|}\binom{\lambda}{\mu} \frac{|\mu|!}{|\lambda|!}\frac{P^{(L,a)}_{\mu} (\mathbf{x})P^{(C)}_{\lambda} (\mathbf{z})} {[a+\xi]_{\mu}^{(\beta)}}.
\end{eqnarray}
Then (\ref{PCtoPL}) can be obtained by comparing the coefficient of $P^{(C)}_{\lambda}(\mathbf{z})$ in both sides of (\ref{PCPLPC}).

\subsection{Constraints of the generalized hypergeometric function $_pF_r(\mathbf{a;b;x})$}
Let us consider the generalized hypergeometric functions with one set of variables.
From (\ref{Wrep2}), it is easy to obtain that
\begin{eqnarray}\label{D-Wcon}
\left(\hat{E}_1-\hat{W}_{1}^{(p)}(\mathbf{a;x})\right){_pF^{(\beta)}_0}(\mathbf{a};\mathbf{x})=0.
\end{eqnarray}

The action of $\hat{O}^{-1}(\mathbf{b};\mathbf{x})$ on (\ref{D-Wcon}) gives
\begin{eqnarray}
\left(\hat{E}_1-\mathfrak{W}_{1}^{(p,r)} \right){_pF^{(\beta)}_r}(\mathbf{a};\mathbf{x})=0,
\end{eqnarray}
where $\mathfrak{W}_{1}^{(p,r)}= \hat{O}^{-1}(\mathbf{b;x})\circ\hat{W}_{1}^{(p)}(\mathbf{a;x})\circ \hat{O}(\mathbf{b;x})$.
However, $\mathfrak{W}_{1}^{(p,r)}$ is not a well-defined differential operator, since it neither belongs to the family of $\hat W$-operators (\ref{Wops}) nor has any specific forms.

In order to obtain constraints for the generalized hypergeometric functions $_pF^{(\beta)}_r(\mathbf{a;b;x})$ with $r>0$, it is necessary to search for other types of constraints.
Due to the relations (\ref{definefpr}), let us consider (\ref{W-W+}) with $(r,n)=(0,1)$ and $\mathbf{z}=1^N$
\begin{eqnarray}\label{Cons}
\hat{W}^{(1)}_{-1}(\xi;\mathbf{x}){_pF^{(\beta)}_0}(\mathbf{a;x})&=&\left(\hat{W}^{(p)}_{1}(\mathbf{a;z}){_p\mathcal{F}^{(\beta)}_0}(\mathbf{a;x;z})\right)\mid_{\mathbf{z}=1^N}\nonumber\\
&=&\hat{O}(\mathbf{a;z})\left(p_1(\mathbf{z}){_0F^{(\beta)}_0(\mathbf{x;z})}\right)|_{\mathbf{z}=1^N}\nonumber\\
&=&\sum_{\lambda}\frac{P^{(C)}_{\lambda}(\mathbf{x})}{|\lambda|!P^{(C)}_{\lambda}(1^N)}\sum_{\mu=\lambda+\Box}\binom{\mu}{\lambda}\prod_{j=1}^{p}[a_j]^{(\beta)}_{\mu}
\frac{P^{(C)}_{\mu}(\mathbf{z})}{|\mu|}|_{\mathbf{z}=1^N}\nonumber\\
&=&\sum_{\lambda}\prod_{j=1}^{p}[a_j]^{(\beta)}_{\lambda}\frac{P^{(C)}_{\lambda}(\mathbf{x})}{|\lambda|!} \mathcal{K}_{\lambda}^{(p)}(\mathbf{a}),
\end{eqnarray}
where
\begin{eqnarray}
\mathcal{K}_{\lambda}^{(p)}(\mathbf{a}):=\sum_{\mu=\lambda+\Box}\binom{\mu}{\lambda}\prod_{j=1}^{p}
\prod_{i=1}^{N}(a_j+\lambda_i+\beta(1-i))\frac{P^{(C)}_{\mu}(1^N)}{|\mu|P^{(C)}_{\lambda}(1^N)},\ \ p\in\mathbb{N}.
\end{eqnarray}
Some examples of $\mathcal{K}_{\lambda}^{(p)}(\mathbf{a})$ were provided in Refs.\cite{Yan,Kaneko,Macdonald13}
\begin{eqnarray}\label{K012}
	\mathcal{K}_{\lambda}^{(0)}&=&N, \nonumber\\
	\mathcal{K}_{\lambda}^{(1)}(a_1)&=&|\lambda|+a_1N, \nonumber\\
	\mathcal{K}_{\lambda}^{(2)}(a_1,a_2)&=&\sum_{(i,j)\in\lambda}[(2j-1)+\beta(N-2i+1)]+(a_1+a_2)|\lambda|+a_1a_2N.
\end{eqnarray}

We define higher circular CS operators
\begin{eqnarray}\label{W0p} \hat{W}_0^{(p)}(\mathbf{a;x})=\mathbf{S}\circ\left(\sum_{j=1}^{N}\prod_{i=1}^{p}(\hat{\pi}_j+a_i)\right)\circ\mathbf{S},
\end{eqnarray}
where $\hat{\pi}_j=x_j\frac{\partial}{\partial x_j}+{\beta\over 2}\sum_{k\neq j}\frac{x_j+x_k}{x_j-x_k}(1-s_{j,k})$
are the Heckman-Dunkl operators \cite{Heckman}.

The first three members of (\ref{W0p}) are
\begin{eqnarray}\label{W012}
\hat{W}_0^{(0)}&=&N,\nonumber\\ \hat{W}_0^{(1)}(a_1;\mathbf{x})&=&\hat{E}_1+a_1N, \nonumber\\
\hat{W}_0^{(2)}(a_1,a_2;\mathbf{x})&=&\hat{D}_2+(a_1+a_2+2-\xi)\hat{E}_1+a_1a_2N.
\end{eqnarray}

We make a conjecture that
\begin{eqnarray}\label{WPKP}
\hat{W}_0^{(p)}(\mathbf{a;x})P_{\lambda}(\mathbf{x})
=\mathcal{K}_{\lambda}^{(p)}(\mathbf{a})P_{\lambda}(\mathbf{x}),\quad \forall p\in \mathbb{N}.
\end{eqnarray}
Here we have checked the cases of $p=0,1,2$ with the help of the formulas (\ref{K012}) and (\ref{W012}).

Based on (\ref{Cons}) and the conjecture (\ref{WPKP}), we have
\begin{eqnarray}\label{W0W-1}
(\hat{W}^{(1)}_{-1}(\xi,\mathbf{x})-\hat{W}_0^{(p)}(\mathbf{a;x})){_pF^{(\beta)}_0}(\mathbf{a;x})=0.
\end{eqnarray}
Then the action of $\hat{O}^{-1}(\mathbf{b};\mathbf{x})$ on (\ref{W0W-1}) gives the constraints
\begin{eqnarray}\label{ConsF}
	\left(\hat{W}_{-1}^{(r+1)}(\xi,\mathbf{b;x})-\hat{W}_0^{(p)}(\mathbf{a;x})\right){_pF^{(\beta)}_r (\mathbf{a;b;x})}=0,\quad r \in \mathbb{N}.
\end{eqnarray}

Especially, we have
\begin{eqnarray}\label{ConsEX}
	\left(\hat{W}_{-1}^{(2)}(\xi,b;\mathbf{x})-\hat{W}_0^{(2)}(a_1,a_2;\mathbf{x})\right){_2F^{(\beta)}_1}(a_1,a_2;b_1;\mathbf{x})&=&0,\nonumber\\
	\left(\hat{W}_{-1}^{(2)}(\xi,b;\mathbf{x})-\hat{W}_0^{(1)}(a_1;\mathbf{x})\right){_1F^{(\beta)}_1}(a_1;b_1;\mathbf{x})&=&0,\nonumber\\
	\left(\hat{W}_{-1}^{(1)}(\xi;\mathbf{x})-\hat{W}_0^{(2)}(a_1,a_2;\mathbf{x})\right){_2F^{(\beta)}_0}(a_1,a_2;\mathbf{x})&=&0.
\end{eqnarray}

By comparing (\ref{ConsEX}) with (\ref{eigenPX}) for $P_{\lambda}^{(J,a,b)}$, $P_{\lambda}^{(L,a)}$ and $P_{\lambda}^{(B,a)}$, we find that the multi-variable Jacobi, Laguerre
and Bessel polynomials with rectangular partition can be represented by the generalized hypergeometric functions
\cite{Hallnas09,Lassalle1,Lassalle2,Lassalle3}
\begin{eqnarray}
	P^{(J,a,b)}_{M^N}(\mathbf{x})&\sim&{_2F^{(\beta)}_1}(-M,M+a+b+\xi;a+\xi;\mathbf{x}),\nonumber\\
	P^{(L,a)}_{M^N}(\mathbf{x})&\sim&{_1F^{(\beta)}_1}(-M;a+\xi;\mathbf{x}),\nonumber\\
	P^{(B,a)}_{M^N}(\mathbf{x})&\sim&{_2F^{(\beta)}_0}(-M,M+a+\xi;-\mathbf{x}/2).
\end{eqnarray}

\section{Integral chains}\label{super}
\subsection{Inner products and orthogonality}
For any symmetric polynomials $f$ and $g$, we define the inner product
\begin{eqnarray}\label{Inner}
	\langle f,g \rangle^{(X)}=
	\int_{D^{(X)} (\mathbf{y})}d\mu^{(X)}(\mathbf{y}) (f(\mathbf{y}))^{\dagger}g(\mathbf{y}),
\end{eqnarray}
where $X\in\{R,C,H,L,J,B\}$, $(f(\mathbf{y}))^\dagger$ is the complex conjugate of $f(\mathbf{y})$, $D^{(X)}(\mathbf{y})$ is the domain of integration and $d\mu^{(X)}(\mathbf{y})$
is the measures which is given by the squared norms of ground wave functions with the transformed variables (the details see Table \ref{Tab3})
\begin{eqnarray}\label{measure}
d\mu^{(X)}(\mathbf{y})&\sim&\left(\Psi_0^{(X)}(\mathbf{y(x)})\right)^{\dagger}\Psi_0^{(X)}(\mathbf{y(x)})\left(\det_{i,j}\left(\frac{\partial y_j}{\partial x_i}
\right)\right)^{-1}d\mathbf{y},\quad \mathbf{y}\in D^{(X)}(\mathbf{y}).
\end{eqnarray}
\begin{table*}[h!]
	\renewcommand\arraystretch{2.5}
	\centering
	\caption{The measures $d\mu^{(X)}(\mathbf{y})/d\mathbf{y}$ \cite{Macdonald,Hallnas09,Lassalle1,Lassalle2,Lassalle3,Diejen} associated with CS models, where $\mathbf{T}_N=\{\mathbf{y}\in\mathbb{C}^N\mid|y_i|=1\} $ and $\Delta(\mathbf{y})=\prod_{1\le j<k\le N}(y_j-y_k)$ is the Vandermonde determinant}\label{Tab3}
	\begin{tabular}{c|c|c|c|c}
		\toprule[2pt]
		Type &$\left(\Psi_0(\mathbf{x})\right)^{\dagger}
		\Psi_0(\mathbf{x})$ & \makecell{Coordinate \\ transformation} & $d\mu^{(X)}(\mathbf{y})/d\mathbf{y}$ & $D^{(X)}(\mathbf{y})$  \\
		\midrule
		$R$ & $\Delta^{2\beta}(\mathbf{x})$ & $y_j=x_j$& $\Delta^{2\beta}(\mathbf{y})$ & $\mathbf{y}\in\mathbb{R}^N$ \\
		$C$ & $\prod_{j<k} \sin^{2\beta} (\frac{\pi}{L}(x_j-x_k))$ & $y_j=e^{{2\pi\mathbf{i}\over L}x_j}$ & $\Delta^{\beta}(\mathbf{y}^{-1})\Delta^{\beta}(\mathbf{y})\prod_{j=1}^{N}y_j^{-1}$ &  $\mathbf{y}\in \mathbf{T}_N$\\
		$H$ & $\Delta^{2\beta} (\mathbf{x})\prod_{j=1}^{N}e^{-{\omega\over 2}x_j^2}$ & $y_j=\sqrt{\omega}x_j$ & $\Delta^{2\beta} (\mathbf{y})\prod_{j=1}^{N}e^{-y_j^2}$ & $\mathbf{y}\in\mathbb{R}^N$\\
		$L$ & $\Delta^{2\beta}(\mathbf{x}^2) \prod_{j=1}^{N}x_j^{2a+1}e^{-\omega x^2_j}$ & $y_j=\omega x_j$ &  $\Delta^{2\beta}(\mathbf{y}) \prod_{j=1}^{N}y_j^ae^{-y_j}$ & $\mathbf{y}\in\mathbb{R}^N_+$\\
		$J$ & \makecell{$\prod_{j=1}^{N} \sin^{2a+1}(x_j)\cos^{2b+1}(x_j)$\\$\times\Delta^{2\beta}(\sin^2 \mathbf{x})$} & $y_j=\sin^2x_j$ & $\Delta^{2\beta}(\mathbf{y})\prod_{j=1}^{N} y_j^a(1-y_j)^b$& $\mathbf{y}\in[0,1]^N$ \\
		$B$ & $\Delta^{2\beta}(e^{\mathbf{x}}) \prod_{j=1}^{N} e^{(a+1)x_j-2e^{-x_j}}$ & $y_j=e^{x_j}$ &  $\Delta^{2\beta}(\mathbf{y})\prod_{j=1}^{N} y_j^{a}e^{-2y_j^{-1}}$ &  $\mathbf{y}\in\mathbb{R}_+^N$\\
		\bottomrule[2pt]
	\end{tabular}
\end{table*}

The CS operators $\mathcal{H}^{(X)}$ are self-adjoint with respect to the corresponding inner products
\begin{eqnarray}\label{self}
	\langle \mathcal{H}^{(X)}f,g \rangle^{(X)}=\langle f,\mathcal{H}^{(X)}g \rangle^{(X)}.
\end{eqnarray}
By comparing $d\mu^{(H)}$ and $d\mu^{(L,a)}$ with $d\mu^{(J,a,b)}$, one can obtain the limits \cite{BakerCMP}
\begin{eqnarray}\label{limits}
	&&\underset{b\rightarrow\infty}{\lim} (-2b)^{|\lambda|} c_{\lambda}^{(H)} P_{\lambda}^{(J,a,b)} ((1-\frac{\mathbf{y}}{b})/2)=P_{\lambda}^{(H)}(\mathbf{y}),\nonumber\\ &&\underset{b\rightarrow\infty}{\lim}b^{|\lambda|}c_{\lambda}^{(L)}P_{\lambda}^{(J,a,b)} (\frac{\mathbf{y}}{b})=P_{\lambda}^{(L,a)}(\mathbf{y}).
\end{eqnarray}

For the case of $X\in\{C,J,B\}$, it is easy to verify that the eigenvalue (\ref{eigenPC}) satisfies the property $\epsilon_{\lambda}^{(X)}\neq\epsilon_{\mu}^{(X)}$ for $\lambda\neq\mu$.
By means of (\ref{eigenPX}) and (\ref{self}), one can prove the symmetric polynomials $P^{(X)}_{\lambda}$ are orthogonal \cite{Macdonald,Hallnas09,Lassalle1,Lassalle2,Lassalle3,Diejen}, i.e.,
\begin{eqnarray}\label{Orth}
	\langle P^{(X)}_{\lambda},P^{(X)}_{\mu}\rangle^{(X)} \propto \delta_{\lambda,\mu}.
\end{eqnarray}
In addition, the orthogonality of Hermite and Laguerre polynomials has been given by \cite{Lassalle1,Lassalle2,Lassalle3,BakerCMP
} which also can be viewed as the property inheriting from the Jacobi case by the limits (\ref{limits}).

\subsection{$\beta$-deformed integrals and superintegrability}
Let us now discuss the partition functions
\begin{eqnarray}\label{Zxt}
Z^{(X)}\{\mathbf{t}\}=\int_{D^{(X)} (\mathbf{x})}d\mu^{(X)}(\mathbf{x})e^{\sum_{k=1}^{\infty}\frac{1}{k}t_k h^{(X)}_k(\mathbf{x})},
\end{eqnarray}
where $h_k^{(X)}(\mathbf{x})=p_k(\mathbf{x})$ for $X\in\{C,H,L,J\}$, $h_k^{(B)}(\mathbf{x})=p_{-k}(\mathbf{x})$ and $\mathbf{t}=(t_1,t_2,\cdots)$.

For later convenience, we denote the corresponding normalized averages
\begin{eqnarray}\label{Average}
\left\langle f(\mathbf{x})\right \rangle^{(X)}_{\mathbf{x}}=\frac{1}{Z^{(X)}}\int_{D^{(X)}(\mathbf{x})}f(\mathbf{x}) d\mu^{(X)}(\mathbf{x})=\frac{\langle1, f\rangle^{(X)}}{\langle1, 1\rangle^{(X)}},
\end{eqnarray}
for any symmetric polynomial $f$, where $Z^{(X)}=Z^{(X)}\{0,0,\cdots\}$.

(i) Circular case

For the circular case, the partition function (\ref{Zxt}) corresponds to the $\beta$-deformation of unitary matrix integral \cite{Macdonald}
\begin{eqnarray}\label{ZCt}
	Z^{(C)}\{\mathbf{t}\}&=&\int_{\mathbf{T}_N} d\mathbf{x}\Delta^{\beta}(\mathbf{x}^{-1}) \Delta^{\beta}(\mathbf{x})\prod_{i=1}^{N}x_i^{-1}e^{\sum_{k=1}^{\infty}\frac{1}{k}t_k\sum_{i=1}^{N}x_i^k}\nonumber\\
	&\sim&\int_{N\times N}[dU]_{\beta} e^{\sum_{k=1}^{\infty}\frac{1}{k}t_kTrU^k}.
\end{eqnarray}

The normalized squared norms of Jack polynomial are given by \cite{Macdonald}
\begin{eqnarray}\label{Jackav}
	\mathcal{N}_{\lambda}^{(C)}=\langle J_{\lambda}(\mathbf{x})J_{\lambda}(\mathbf{x}^{-1}) \rangle^{(C)}_{\mathbf{x}} =\frac{[N\beta]^{(\beta)}_{\lambda}}{[\xi]^{(\beta)}_{\lambda}}j_{\lambda},
\end{eqnarray}
which can also be proved by comparing both sides of the following equation
\begin{eqnarray}
	J_{\mu}(\mathbf{z})&=&\langle_1\mathcal{F}_0 ^{(\beta)}(\xi;\mathbf{x};\mathbf{z})J_{\mu}(\mathbf{z})\rangle^{(C)}_{\mathbf{x}}\nonumber\\
	&=&\langle_1\mathcal{F}_0^{(\beta)}(\xi;\mathbf{x};\mathbf{z}) J_{\mu}(\mathbf{x}^{-1}) \rangle^{(C)}_{\mathbf{x}}\nonumber \\
	&=& \sum_{\lambda}\frac{[\xi]^{(\beta)}_{\lambda}} {[N\beta]^{(\beta)}_{\lambda}}\langle J_{\lambda}(\mathbf{x})J_{\mu}(\mathbf{x}^{-1}) \rangle^{(C)}_{\mathbf{x}}\frac{J_{\lambda}(\mathbf{z})}{j_{\lambda}},
\end{eqnarray}
where we have used (\ref{fzfx}) and orthogonality of Jack polynomials (\ref{Orth}).

Let us consider the following $\beta$-deformed matrix models
\begin{eqnarray}\label{CirZx}
	Z_k(\mathbf{y})&=&\int_{N\times N}[dV]_{\beta} e^{Tr(V^{\dagger})^k+TrVY}\nonumber\\
	&=&\int_{N\times N}[dU]_{\beta}I_{IZ}(U,Y) e^{Tr(U^{\dagger})^k},
\end{eqnarray}
where $k\in\mathbb{Z}_+$ and $Y=\mathbf{diag}\{y_1,\cdots,y_N\}$.

By means of (\ref{ZCt}) and (\ref{Jackav}), we obtain the eigenvalue integrals and character expansions of (\ref{CirZx})
\begin{eqnarray}\label{CirZx2}
Z_k(\mathbf{y})&\sim&\int_{\mathbf{T}_N} d\mathbf{x} \Delta^{\beta}(\mathbf{x}^{-1}) \Delta^{\beta}(\mathbf{x})\prod_{i=1}^{N}x_i^{-1}
{_0\mathcal{F}^{(\beta)}_0}(\mathbf{x};\mathbf{y})e^{p_k(\mathbf{x})}  \nonumber\\
&=&\sum_{\lambda}\frac{1}{[\xi]_{\lambda} ^{(\beta)}}\frac{J_{\lambda}\{\beta^{-1}\delta_{n,k}\}J_{\lambda}(\mathbf{y})}{j_{\lambda}},\nonumber\\
&=&{_0 F_1^{(\beta)}}(\xi;\mathbf{y}^k).
\end{eqnarray}
When $k=1$ in (\ref{CirZx2}), it gives the partition function of $\beta$-deformed Brezin-Gross-Witten (BGW) model with character phase \cite{MirBGW}.

(ii) Hermite and Laguerre cases

For the Hermite and Laguerre cases, the partition function (\ref{Zxt}) corresponds to the $\beta$-deformed Hermite and Laguerre ensembles, respectively\cite{Dumitriu}
\begin{subequations}\label{ZHLt}
\begin{eqnarray}
	Z^{(H)}\{\mathbf{t}\}&=&\int_{\mathbb{R}^N}d\mathbf{x} \Delta^{2\beta}(\mathbf{x}) e^{-p_2(\mathbf{x})}e^{\sum_{k=1}^{\infty}\frac{1}{k}t_k \sum_{i=1}^{N}x_i^k},\\
	Z^{(L,a)}\{\mathbf{t}\}&=&\int_{\mathbb{R}_+^N}d\mathbf{x}\Delta^{2\beta}(\mathbf{x})\prod_{j=1}^{N}x_j^a e^{-p_1(\mathbf{x})}e^{\sum_{k=1}^{\infty}\frac{1}{k}t_k \sum_{i=1}^{N}x_i^k}\label{intL}.
\end{eqnarray}
\end{subequations}

The superintegrability relations are  \cite{Cassia,MirSum}
\begin{subequations}\label{JackavHL}
\begin{eqnarray}
	&&\langle J_{\lambda}(\mathbf{x})\rangle^{(H)} _\mathbf{x}= 2^{-\lambda}[N\beta]^{(\beta)}_{\lambda}J_{\lambda}\{\beta^{-1}\delta_{n,2}\} = \frac{J_{\lambda}\{\beta^{-1}\delta_{n,2}\}J_{\lambda}\{N\}}{J_{\lambda}\{2\beta^{-1}\delta_{n,1}\}},\\
	&&\langle J_{\lambda}(\mathbf{x}) \rangle^{(L,a)} _\mathbf{x}=[a+\xi]_{\lambda}^{(\beta)}[N\beta]^{(\beta)}_{\lambda}J_{\lambda}\{\beta^{-1}\delta_{n,1}\} = \frac{J_{\lambda}\{\beta^{-1}(a+\xi)\}J_{\lambda}\{N\}}{J_{\lambda}\{\beta^{-1}\delta_{n,1}\}},
\end{eqnarray}	
\end{subequations}
where
\begin{eqnarray}
	[a]_{\lambda}^{(\beta)}=\prod_{(i,j)\in\lambda} [a+(j-1)+\beta(1-i)]=\frac{J_{\lambda}\{a\beta^{-1}\}}{J_{\lambda}\{\beta^{-1}\delta_{n,1}\}}.
\end{eqnarray}
The relations (\ref{JackavHL}) can also be proved by means of $W$-representations \cite{Rui22}.

Let us now give a new proof of (\ref{JackavHL}).
Using the expansion formulas (\ref{expansion1}) and (\ref{expansion2}), we obtain the squared norms of Hermite and Laguerre polynomials \cite{Lassalle1,Lassalle2,Lassalle3,BakerCMP}
\begin{subequations}
\begin{eqnarray}
	&&\mathcal{N}_{\lambda}^{(H)}=\langle P^{(H)}_{\lambda}(\mathbf{x}) P^{(H)}_{\lambda}(\mathbf{x}) \rangle^{(H)}_{\mathbf{x}} =\frac{2^{|\lambda|}|\lambda|!}{P^{(C)}_{\lambda}(1^N)},\\
	&&\mathcal{N}_{\lambda}^{(L,a)}=\langle P^{(L,a)}_{\lambda}(\mathbf{x}) P^{(L,a)}_{\lambda}(\mathbf{x})\rangle^{(L,a)}_{\mathbf{x}}=\frac{[a+\xi]_{\lambda}^{(\beta)}}{P^{(C)}_{\lambda}(1^N)|\lambda|!}.
\end{eqnarray}
\end{subequations}
Then, by the expansion formulas (\ref{p2p1F}), we recover the formulas in Ref.\cite{BakerCMP}
\begin{subequations}\label{avFP}
\begin{eqnarray}
\langle_0\mathcal{F}_0^{(\beta)}(\mathbf{x};\mathbf{z})e^{-p_2(\mathbf{z}/2)}P_{\mu}^{(H)}(\mathbf{x})
\rangle^{(H)}_{\mathbf{x}}&=&\frac{P^{(C)}_{\mu}(\mathbf{z})}{P^{(C)}_{\mu}(1^N)},\\
\langle_0\mathcal{F}_1^{(\beta)}(a+\xi;\mathbf{x};-\mathbf{z})e^{p_1(\mathbf{z})}P_{\mu}^{(L,a)}(\mathbf{x})\rangle^{(L,a)}_\mathbf{x}
&=& \frac{P^{(C)}_{\mu}(\mathbf{z})} {P^{(C)}_{\mu}(1^N)|\mu|!}.
\end{eqnarray}
\end{subequations}
With the help of the expansions of the generalized hypergeometric functions (\ref{defHyper}) and the Cauchy identities (\ref{Cauchy}), the averages (\ref{JackavHL})
can be derived by taking the partition $\mu=\emptyset$ in (\ref{avFP}).

Moreover, the averages of generalized hypergeometric functions are given by
\begin{subequations}
\begin{eqnarray}
	&&\langle_p\mathcal{F}_r^{(\beta)} (a_1,\cdots,a_p;b_1,\cdots,b_r;\mathbf{x;z})\rangle^{(H)}_{\mathbf{x}} \nonumber\\ &=&{_{p}F_{r}^{(\beta)}}(a_1,\cdots,a_p;b_1,\cdots,b_r;\mathbf{z}^2/4),\\
	&&\langle _p\mathcal{F}_r^{(\beta)} (a_1,\cdots,a_p;b_1,\cdots,b_r;\mathbf{x;z})\rangle^{(L,a)}_{\mathbf{x}} \nonumber\\ &=&{_{p+1}F_{r}^{(\beta)}}(a+\xi,a_1,\cdots, a_p;b_1,\cdots,b_r;\mathbf{z}).
\end{eqnarray}
\end{subequations}
Especially, taking $t_k=\sum_{j=1}^{N}z_j^k$ for $k\in\mathbb{Z}_+$ in (\ref{ZHLt}), one can obtain
\begin{subequations}
\begin{eqnarray}
	\frac{Z^{(H)}\{\mathbf{t}\}}{Z^{(H)}}&=&\langle _1\mathcal{F}_0^{(\beta)}(N\beta;\mathbf{x;z}) \rangle^{(H)}_{\mathbf{x}}\nonumber\\
	&=&{_{1}F_{0}^{(\beta)}}(N\beta;\mathbf{z}^2/4).\\
	\frac{Z^{(L,a)}\{\mathbf{t}\}}{Z^{(L,a)}}&=&\langle _1\mathcal{F}_0^{(\beta)}(N\beta;\mathbf{x;z}) \rangle^{(L,a)}_{\mathbf{x}}\nonumber\\
	&=&{_{2}F_{0}^{(\beta)}}(a+\xi,N\beta;\mathbf{z}).
\end{eqnarray}
\end{subequations}

(iii) Jacobi case

For the Jacobi case, the partition function (\ref{Zxt}) corresponds to the $\beta$-deformed Jacobi ensemble \cite{Lassalle1,Lassalle2,Lassalle3,Diejen}
\begin{eqnarray}\label{ZJt}
	Z^{(J,a,b)}\{\mathbf{t}\}= \int_{[0,1]^N}d\mathbf{x} \Delta^{2\beta}(\mathbf{x})\prod_{j=1}^{N} x_j^a(1-x_j)^be^{\sum_{k=1}^{\infty}\frac{1}{k}t_k \sum_{i=1}^{N}x_i^k},
\end{eqnarray}
and the superintegrability relation is \cite{Kadell}
\begin{eqnarray}\label{JackavJ}
	\langle J_{\lambda}(\mathbf{x}) \rangle^{(J,a,b)}_{\mathbf{x}} =\frac{J_{\lambda}\{N\}J_{\lambda}\{\beta^{-1}(a+\xi)\}}{J_{\lambda}\{\beta^{-1}(a+b+2\xi)\}}.
\end{eqnarray}
Note that the averages (\ref{JackavHL}) can also be obtained from (\ref{JackavJ}) by the limits (\ref{limits}).

It follows from (\ref{JackavJ}) that
\begin{eqnarray}
	&&\langle _p\mathcal{F}_r^{(\beta)} (a_1,\cdots,a_p;b_1,\cdots,b_r;\mathbf{x;z})\rangle^{(J,a,b)}_{\mathbf{x}} \nonumber\\ &=&{_{p+1}F_{r+1}^{(\beta)}}(a+\xi,a_1,\cdots,a_p;a+b+2\xi,b_1,\cdots,b_r; \mathbf{z}).
\end{eqnarray}
Especially, taking $t_k=\sum_{j=1}^{N}z_j^k$ for $k\in\mathbb{Z}_+$ in (\ref{ZJt}), one can obtain
\begin{eqnarray}
	\frac{Z^{(J,a,b)}\{\mathbf{t}\}}{Z^{(J,a,b)}}&=&\langle _1\mathcal{F}_0^{(\beta)} (N\beta;\mathbf{x;z}) \rangle^{(J,a,b)}_{\mathbf{x}}\nonumber\\
	&=&{_{2}F_{1}^{(\beta)}}(a+\xi,N\beta;a+b+2\xi;\mathbf{z}).
\end{eqnarray}

(iv) Bessel case

For the Bessel case, the partition function (\ref{Zxt}) becomes
\begin{eqnarray}\label{BessInt}
	Z^{(B,a)}\{\mathbf{t}\}&=&\int_{\mathbb{R}_+^N}  d\mathbf{x}\Delta^{2\beta}(\mathbf{x})\prod_{j=1}^{N}x_j^ae^{-2p_{1}(\mathbf{x}^{-1})} e^{\sum_{k=1}^{\infty}\frac{1}{k}t_k \sum_{i=1}^{N}x_i^{-k}}\nonumber\\
	&\sim&\int_{\mathbb{R}_+^N}  d\mathbf{x}\Delta^{2\beta}(\mathbf{x})\prod_{j=1}^{N}x_j^{-a-2\xi}e^{-p_{1}(\mathbf{x})} e^{\sum_{k=1}^{\infty} \frac{2^{-k}}{k}t_k \sum_{i=1}^{N}x_i^k}.
\end{eqnarray}

Comparing the integral (\ref{BessInt}) with the Laguerre case (\ref{intL}), we obtain superintegrability relation
\begin{eqnarray}\label{JackavB}
	\langle J_{\lambda}(\mathbf{x}^{-1})\rangle^{(B,a)} _\mathbf{x}&=&2^{-|\lambda|}\langle J_{\lambda}(\mathbf{x}) \rangle^{(L,-a-2\xi)}_\mathbf{x}\nonumber\\
	&=&\frac{J_{\lambda}\{N\}J_{\lambda}\{-\beta^{-1}(a+\xi)\}}{J_{\lambda}\{2\beta^{-1}\delta_{n,1}\}}.
\end{eqnarray}
It follows from (\ref{JackavB}) that
\begin{eqnarray}
	&&\langle _p\mathcal{F}_r^{(\beta)} (a_1,\cdots,a_p;b_1,\cdots,b_r;2\mathbf{x^{-1}};\mathbf{z})\rangle^{(B,a)}_{\mathbf{x}} \nonumber\\ &=&{_{p+1}F_{r}^{(\beta)}}(-a-\xi,a_1,\cdots,a_p;b_1,\cdots,b_r; \mathbf{z}).
\end{eqnarray}
Especially, taking $t_k=\sum_{j=1}^{N}z_j^k$ for $k\in\mathbb{Z}_+$ in (\ref{BessInt}), we have
\begin{eqnarray}
	\frac{Z^{(B,a)}\{\mathbf{t}\}}{Z^{(B,a)}}&=&\langle _1\mathcal{F}_0^{(\beta)}(N\beta;\mathbf{x^{-1};z}) \rangle^{(B,a)}_{\mathbf{x}}\nonumber\\
	&=&{_2F_0^{(\beta)}}(-a-\xi,N\beta;\mathbf{z}/2).
\end{eqnarray}

\subsection{Integral chains and superintegrability}
A hierarchy of $\beta$-deformed partition functions with $W$-representations was constructed in Refs.\cite{Rui22,MirInter,MirSkew}. For the case of $\beta=1$, these partition functions can be expressed as integral
chains \cite{Orlov,Alexandrovchain,Alexandrov14} , where the integral chains are certain chain-like structures constructed by individual integrals.
Note that some $\beta$-deformed partition functions \cite{Oreshina03,Oreshina04} can be constructed in terms of the $\beta$-deformed HCIZ integral \cite{Dumitriu}.
However, for the general $\beta$-deformed partition functions in the hierarchy, their integral representations are still unknown. In this subsection, we will construct certain desired integral representations.

With the help of the measures $d\mu^{(C)}$ and $d\mu^{(L,a)}$, we introduce the integral with two set of variables
\begin{eqnarray}\label{INa}
	I_N^{(a)}&=&\int_{\mathbf{T}_{N}}d\mu^{(C)}(\mathbf{y})\int_{\mathbb{R}_{+}^N}d\mu^{(L,a)}(\mathbf{x})e^{p_1(\mathbf{x})} {_0\mathcal{F}^{(\beta)}_0 (\mathbf{-x};\mathbf{y})} \prod_{j=1}^{N}y_j^{a+\xi}\nonumber\\
	&=&\int_{\mathbf{T}_{N}}d\mathbf{y} \int_{\mathbb{R}_{+}^N}d\mathbf{x}\Delta^{2\beta}(\mathbf{x}) \Delta^{2\beta}(\mathbf{y}) {_0\mathcal{F}_0^{(\beta)}} (\mathbf{-x;y})\prod_{j=1}^{N}x_j^{a}y_j^{a}.
\end{eqnarray}
Due to the relation (\ref{F00IZ}), the inserted function ${_0\mathcal{F}_0^{(\beta)}}(\mathbf{-x;y})$ is the $\beta$-HCIZ integral in essence. Thus when specialized to the $a=0$ case in (\ref{INa}),
it reproduces the integral in Ref.\cite{Oreshina04}.

For the integral (\ref{INa}), the normalized average is
\begin{eqnarray}
	\langle f(\mathbf{x,y})\rangle^{(a)}_{\mathbf{x,y}} &=&\frac{1}{I_N^{(a)}}\int_{\mathbf{T}_{N}}d\mathbf{y} \int_{\mathbb{R}_{+}^N}d\mathbf{x} \Delta^{2\beta} (\mathbf{x})\Delta^{2\beta}(\mathbf{y})\nonumber\\ &&\times{_0\mathcal{F}^{(\beta)}_0}(\mathbf{-x;y})	 f(\mathbf{x,y})\prod_{j=1}^{N}x_j^{a}y_j^{a},
	\nonumber \\
	&=&\left\langle\left\langle f(\mathbf{x,y}) e^{p_1(\mathbf{x})}{_0\mathcal{F}^{(\beta)}_0 (\mathbf{-x};\mathbf{y})} \prod_{j=1}^{N}y_j^{a+\xi} \right\rangle^{(L,a)}_{\mathbf{x}} \right\rangle^{(C)}_{\mathbf{y}}.
\end{eqnarray}
Due to the Sokhotski theorem \cite{Oreshina04},  we see that the values of $I_N^{(a)}$ and $\langle f(\mathbf{x,y})\rangle^{(a)}_{\mathbf{x,y}}$
do not change when replacing the region of integration $|y_j|=1$ by $y_j\in\mathbf{i}\mathbb{R}$.

Using the formulas (\ref{expansion2}), (\ref{definebino}) and (\ref{PCtoPL}),
we can recover the generalized Laplace transformation of Jack polynomials \cite{Macdonald13,BakerCMP}
\begin{eqnarray}\label{Laplace}
	&&\left\langle J_{\lambda}(\mathbf{x})e^{p_1(\mathbf{x})} {_0\mathcal{F}_0^{(\beta)}(\mathbf{-x};\mathbf{y})} \prod_{j=1}^{N}y_j^{a+\xi}\right\rangle_{\mathbf{x}}^{(L,a)}\nonumber\\
	&=&\left\langle J_{\lambda}(\mathbf{x})\sum_{\mu} P_{\mu}^{(L,a)}(\mathbf{x})P^{(C)}_\mu(1-\mathbf{y}^{-1}) \right\rangle_{\mathbf{x}}^{(L,a)} \nonumber\\
	&=&J_{\lambda}(1^N)\sum_{\nu\subseteq\lambda}(-1)^{|\nu|}\frac{[a+\xi]_{\lambda}^{(\beta)}}{[a+\xi]_{\nu}^{(\beta)}}\binom{\lambda}{\nu}|\nu|! \sum_{\mu}\langle P_{\nu}^{(L,a)}(\mathbf{x}) P_{\mu}^{(L,a)}(\mathbf{x})\rangle_{\mathbf{x}}^{(L,a)} P^{(C)}_{\mu}(1-\mathbf{y}^{-1}) \nonumber\\
	&=& [a+\xi]_{\lambda}^{(\beta)}J_{\lambda}(\mathbf{y}^{-1}).
\end{eqnarray}

By means of (\ref{Jackav}) and (\ref{Laplace}), we obtain the orthogonality condition associated with the integral $I^{(a)}_N$,
\begin{eqnarray}\label{JJxy}
	\langle J_{\lambda}(\mathbf{x})J_{\mu}(\mathbf{y}) \rangle^{(a)}_{\mathbf{x,y}}
	&=&[a+\xi]_{\lambda}^{(\beta)}\langle J_{\lambda}(\mathbf{y}^{-1})J_{\mu}(\mathbf{y}) \rangle^{(C)}_{\mathbf{y}} \nonumber\\
	&=&\mathcal{J}^{(\beta)}_{\lambda}(N,a)j_{\lambda}\delta_{\lambda,\mu},
\end{eqnarray}
where
\begin{eqnarray}
\mathcal{J}^{(\beta)}_{\lambda}(N,a):=\frac{[a+\xi]_{\lambda}^{(\beta)}[N\beta]_{\lambda}^{(\beta)}}{[\xi]_{\lambda}^{(\beta)}}=
\frac{J_{\lambda}\{N\}J_{\lambda}\{\beta^{-1}(a+\xi)\}}{J_{\lambda}\{\beta^{-1}\delta_{n,1}\}J_{\lambda}\{\beta^{-1}\xi\}}.
\end{eqnarray}

Let us construct the following integral chain
\begin{eqnarray}\label{chain}
&&Z_{chain}^{(\beta)}(\mathbf{x,y;a;N};L)\nonumber\\
&=&\left\langle\cdots\left\langle \prod_{l=0}^{L}\prod_{j=1}^{N_l}\prod_{k=1}^{N_{l+1}}(1-x_j^{(l)}y_k^{(l+1)})^{-\beta}\right\rangle_{\mathbf{x}^{(1)},\mathbf{y}^{(1)}}
^{(a_1)}\cdots\right\rangle_{\mathbf{x}^{(L)},\mathbf{y}^{(L)}}^{(a_L)}\nonumber\\
&=&\prod_{l=1}^{L}\frac{1}{I_{N_l}^{(a)}}\int_{\mathbf{T}_{N_l}}d\mathbf{y}^{(l)} \int_{\mathbb{R}_{+}^{N_l}}d\mathbf{x}^{(l)} \Delta^{2\beta} (\mathbf{x}^{(l)})\Delta^{2\beta} (\mathbf{y}^{(l)}){_0\mathcal{F}^{(\beta)}_0}(-\mathbf{x}^{(l)};\mathbf{y}^{(l)})\prod_{j=1}^{N_l} (x_j^{(l)}y_j^{(l)})^{a_l}\nonumber\\ &&\times \prod_{l'=0}^{L}\prod_{j=1}^{N_{l'}}\prod_{k=1}^{N_{l'+1}}(1-x_j^{(l')}y_k^{(l'+1)})^{-\beta} \nonumber\\
&=&\sum_{\lambda}\prod_{l=1}^{L} \mathcal{J}^{(\beta)}_{\lambda}(N_l,a_l) \frac{J_{\lambda}(\mathbf{x})J_{\lambda}(\mathbf{y})}{j_{\lambda}},
\end{eqnarray}
where $L\in\mathbb{N}$, $N_0=N_{L+1}=N$, $\mathbf{N}=(N_1,\cdots,N_L)$, $\mathbf{a}=(a_1,\cdots,a_L)$, $x^{(0)}_j=x_j$, $y^{(L+1)}_j=y_j$ for $j=1,2,\cdots,N$,
$(\mathbf{x}^{(l)})_j=x^{(l)}_j$, $(\mathbf{y}^{(l)})_j=y^{(l)}_j$ for $j=1,\cdots,N_l$ and $l=1,\cdots,L$.
Here, $Z_{chain}^{(\beta)}(\mathbf{x,y;a;N};0) =\prod_{j,k=1}^{N}(1-x_jy_k)^{-\beta}$. In addition, the orthogonality condition (\ref{JJxy}) has been used in the last line of (\ref{chain}).

For any symmetric polynomials $f$, let us denote
\begin{eqnarray}
	\widetilde{\langle f(\mathbf{x}^{(L)}) \rangle}_L=\left\langle\cdots\left\langle \prod_{l=0}^{L-1}\prod_{j=1}^{N_l}\prod_{k=1}^{N_{l+1}}(1-x_j^{(l)}y_k^{(l+1)})^{-\beta}f(\mathbf{x}^{(L)}) \right\rangle_{\mathbf{x}^{(1)},\mathbf{y}^{(1)}}^{(a_1)}\cdots \right\rangle_{\mathbf{x}^{(L)},\mathbf{y}^{(L)}}^{(a_L)}.
\end{eqnarray}
From (\ref{chain}), we obtain the superintegrability relation for the correlator
\begin{eqnarray}\label{tildJA}
\widetilde{\langle J_{\lambda}(\mathbf{x}^{(L)}) \rangle}_L
=\prod_{l=1}^{L}\mathcal{J}^{(\beta)}_{\lambda}(N_l,a_l) J_{\lambda}(\mathbf{x}).
\end{eqnarray}

By the character expansion formula (\ref{chain}), we have
\begin{eqnarray}\label{chhyp}
	Z_{chain}^{(\beta)}(\mathbf{x,y;a;N};L)&=& {_{2L+1}\mathcal{F}_L^{(\beta)}} (N\beta,\mathbf{N} \beta,\mathbf{a}+\Xi;\Xi;\mathbf{x;y}),
\end{eqnarray}
where $\Xi=(\xi_1,\cdots,\xi_L)$ and $\xi_l=(N_l-1)\beta+1$.

By (\ref{eigenWF}), we give the constraints
\begin{eqnarray}\label{chaincon}
\left(\hat{W}^{(2L+1)}_{n}(N\beta,\mathbf{N}\beta,\mathbf{a} +\Xi;\mathbf{x}) -\hat{W}^{(L+1)}_{-n}
(\xi,\Xi;\mathbf{y})\right)Z_{chain}^{(\beta)}(\mathbf{x,y;a;N};L)=0.
\end{eqnarray}

We now turn to discuss two particular cases of (\ref{chain}).

(i) When $\mathbf{a}=\mathbf{0}$ in (\ref{chain}), it gives the $\beta$-deformed partition functions with $W$-representation
\begin{eqnarray}\label{chain1}
	&&Z_{chain}^{(\beta)}(\mathbf{x,y;0;N};L) \nonumber\\
	&=&\prod_{l=1}^{L}\frac{1}{I_{N_l}^{(0)}}\int_{\mathbf{T}_{N_l}}d\mathbf{y}^{(l)} \int_{\mathbb{R}_{+}^{N_l}}d\mathbf{x}^{(l)} \Delta^{2\beta} (\mathbf{x}^{(l)})\Delta^{2\beta} (\mathbf{y}^{(l)}){_0\mathcal{F}^{(\beta)}_0}(-\mathbf{x}^{(l)};\mathbf{y}^{(l)})\nonumber\\
	&&\times \prod_{l'=0}^{L}\prod_{j=1}^{N_{l'}} \prod_{k=1}^{N_{l'+1}}(1-x_j^{(l')}y_k^{(l'+1)})^{-\beta} \nonumber\\
	&=&{_{L+1}\mathcal{F}_0^{(\beta)}}(N\beta, \mathbf{N}\beta;\mathbf{x;y})\nonumber\\
	&=&\exp\left(\beta\sum_{n=1}^{\infty}\frac{W^{(L)}_n(\mathbf{N}\beta;\mathbf{x})p_n(\mathbf{y})}{n}\right)\cdot 1.
\end{eqnarray}
We see that the last line of (\ref{chain1}) reproduces the $W$-representation constructed in Refs.\cite{MirSkew,MirCom}. In addition, the constraints (\ref{chaincon}) become
\begin{eqnarray}
	&&\left(\hat{W}^{(L+1)}_{n}(N\beta,\mathbf{N}\beta;\mathbf{x})-\hat{W}^{(1)}_{-n}
	(\xi;\mathbf{y})\right)Z_{chain}^{(\beta)}(\mathbf{x,y;0;N};L)=0.
\end{eqnarray}

(ii) When $\beta=1$ in (\ref{chain}), it gives the partition functions with $W$-representation
\begin{eqnarray}\label{chain2}
	&&Z_{chain}^{(1)}(\mathbf{x,y;a;N};L) \nonumber\\
	&=&\prod_{l=1}^{L}\frac{1}{I_{N_l}^{(a)}|_{\beta=1}}\int_{\mathbf{T}_{N_l}}d\mathbf{y}^{(l)} \int_{\mathbb{R}_{+}^{N_l}}d\mathbf{x}^{(l)} \Delta^2 (\mathbf{x}^{(l)})\Delta^2 (\mathbf{y}^{(l)}){_0\mathcal{F}^{(1)}_0}(-\mathbf{x}^{(l)};\mathbf{y}^{(l)})\prod_{j=1}^{N_l} (x_j^{(l)}y_j^{(l)})^{a_l}\nonumber\\
	&&\times \prod_{l'=0}^{L}\prod_{j=1}^{N_{l'}} \prod_{k=1}^{N_{l'+1}}(1-x_j^{(l')}y_k^{(l'+1)})^{-1} \nonumber\\
	&=&{_{L+1}\mathcal{F}_0^{(1)}}(N, \mathbf{a+N};\mathbf{x;y})\nonumber\\
	&=&\exp\left(\sum_{n=1}^{\infty}\frac{\mathbb{W}^{(L)}_n(\mathbf{a+N};\mathbf{x})p_n(\mathbf{y})}{n}\right)\cdot 1,
\end{eqnarray}
where $\mathbb{\hat{W}}^{(L)}_{\pm k}=\hat{W}^{(L)}_{\pm k}\mid_{\beta=1}$. We also see that the last line of (\ref{chain2}) reproduces the $W$-representation
constructed in Refs.\cite{MirSkew,MirCom}. Similarly, the constraints (\ref{chaincon}) become
\begin{eqnarray}
&&\left(\mathbb{\hat{W}}^{(L+1)}_{n}(N,\mathbf{a+N};\mathbf{x})-\mathbb{\hat{W}}^{(1)}_{-n}(N;\mathbf{y})\right)Z_{chain}^{(1)}(\mathbf{x,y;a;N};L)=0.
\end{eqnarray}

For the matrix integral \cite{Alexandrov14}
\begin{eqnarray}\label{zchain3}
	&&Z_{(2,l)}\{N_1,N_2,\cdots,N_l|\mathbf{t},\mathbf{\bar{t}}\}\nonumber\\
	&=&\int d\mathcal{K}_{N_1}(X_1|\mathbf{t}) \frac{\prod_{i=2} ^{n-1} e^{i\operatorname{Tr} X_iY_i}dX_idY_i} {\prod_{i=1}^{n-1}\det \left(I_{N_i}\otimes I_{N_{i+1}}-X_i\otimes Y_{i+1}\right)}d\mathcal{K}_{N_n}(Y_n|\bar{\mathbf{t}}),
\end{eqnarray}
where $d\mathcal{K}_N(Y|\mathbf{t})=dY\int_{N\times N}e^{\sum_n\frac{1}{n}t_n\mathrm{Tr}X^n}e^{i\mathrm{Tr}XY}dX$, it is the hypergeometric $\tau$-function of 2-component
KP hierarchy as well as the generating function of the weighted Hurwitz numbers \cite{Alexandrov24,Alexandrov14}. The partition functions (\ref{chain2}) with $\mathbf{a}=\mathbf{0}$
and (\ref{zchain3}) are equivalent, since they have the same character expansions.

Similar to (\ref{JJxy}), we have
\begin{eqnarray}\label{skewJJ}
	&&\langle J_{\lambda}(\mathbf{x})J_{\mu}(\mathbf{y}) {_0\mathcal{F}^{(\beta)}_0(\mathbf{y,z})}\rangle^{(a)}_{\mathbf{x,y}}\nonumber\\
	&=&[a+\xi]_{\lambda}^{(\beta)}\langle J_{\lambda}(\mathbf{y}^{-1})J_{\mu}(\mathbf{y}){_0\mathcal{F}^{(\beta)}_0(\mathbf{y,z})} \rangle^{(C)}_{\mathbf{y}} \nonumber\\
	&=&\sum_{\nu}\frac{j_{\lambda}}{j_{\nu}}\frac{\mathcal{J}^{(\beta)}_{\lambda}(N,a)}{[N\beta]_{\nu}^{(\beta)}}C_{\mu\nu}^{\lambda}J_{\nu}(\mathbf{z}),
\end{eqnarray}
where the coefficient $C_{\mu\nu}^{\lambda}$ is determined by $J_{\mu}(\mathbf{y})J_{\nu}(\mathbf{y}) = \sum_{\lambda}C_{\mu\nu}^{\lambda}J_{\lambda}(\mathbf{y})$.

Then we construct the integral chain
\begin{eqnarray}\label{skewchain}
	&&\bar{Z}^{(\beta)}_{chain}(\mathbf{x,y,z;a;N};L)\nonumber\\
	&=&\left\langle Z_{chain}^{(\beta)}(\mathbf{x},\mathbf{x}^{(L)};\mathbf{\tilde{a};\tilde{N}};L-1)\prod_{j,k=1}^{N_{L}}(1-y_j^{(L)}y_k)^{-\beta}S(\mathbf{y}^{(L)},\mathbf{z};L;\beta) \right\rangle_{\mathbf{x}^{(L)}, \mathbf{y}^{(L)}}^{(a_L)}\nonumber\\
	&=&\frac{1}{I_{N_L}^{(a_L)}}\int_{\mathbf{T}_{N_L}}d\mathbf{y}^{(L)} \int_{\mathbb{R}_{+}^{N_L}} d\mathbf{x}^{(L)} \Delta^{2\beta} (\mathbf{x}^{(L)})\Delta^{2\beta} (\mathbf{y}^{(L)}){_0\mathcal{F}^{(\beta)}_0}(-\mathbf{x}^{(L)};\mathbf{y}^{(L)})\prod_{j=1}^{N_L} (x_j^{(L)}y_j^{(L)})^{a_L}\nonumber\\
	&&\times Z_{chain}^{(\beta)}(\mathbf{x},\mathbf{x}^{(L)} ;\mathbf{\tilde{a};\tilde{N}};L-1)\prod_{j,k=1}^{N_{L}}(1-y_j^{(L)}y_k)^{-\beta}S(\mathbf{y}^{(L)},\mathbf{z};L;\beta)\nonumber\\
	&=&\sum_{\lambda,\nu}\prod_{l=1}^{L}\frac{\mathcal{J}^{(\beta)}_{\lambda}(N_l,a_l)}{\mathcal{J}^{(\beta)}_{\nu}(N_l,a_l)}\frac{J_{\lambda/\nu}(\mathbf{y})J_{\lambda}(\mathbf{x})J_{\nu}(\mathbf{z})}{j_{\lambda}},
\end{eqnarray}
where $\mathbf{\tilde{N}}=(N_1,\cdots,N_{L-1})$, $\mathbf{\tilde{a}}=(a_1,\cdots,a_{L-1})$, $(\mathbf{x})_j=x_j$, $(\mathbf{y})_j=y_j$, $(\mathbf{z})_j=z_j$ for $j=1,\cdots,N_L$ and $S(\mathbf{y}^{(L)},\mathbf{z};L;\beta)= {_{L}\mathcal{F}^{(\beta)}_{2L-1}}(\Xi;\mathbf{a}+\Xi,\mathbf{\tilde{N}}\beta;\mathbf{y}^{(L)};\mathbf{z})$. Here, the skew Jack polynomial
$J_{\lambda/\nu}(\mathbf{y})=\sum_{\mu}\frac{j_{\lambda}}{j_{\mu}j_{\nu}}C_{\mu\nu}^{\lambda}J_{\mu}(\mathbf{y})$ and (\ref{skewJJ}) have been used in the last line of (\ref{skewchain}).

Similarly, for any symmetric polynomials $f$, let us denote
\begin{eqnarray}
\widetilde{\langle f(\mathbf{y}^{(1)})\rangle}^{skew}_L&=&
\left\langle\left\langle f(\mathbf{y}^{(1)}) Z_{chain}^{(\beta)}(\mathbf{x}^{(1)} ,\mathbf{x}^{(L)};\mathbf{\tilde{a}';\tilde{N}'};L-2)\right\rangle_{\mathbf{x}^{(1)}, \mathbf{y}^{(1)}}^{(a_1)}\right.\nonumber\\
&&\cdot\left.\prod_{j,k=1}^{N_{L}}(1-y_j^{(L)}y_k)^{-\beta}S(\mathbf{y}^{(L)},\mathbf{z};L;\beta) \right\rangle_{\mathbf{x}^{(L)}, \mathbf{y}^{(L)}}^{(a_L)},
\end{eqnarray}
where $\mathbf{\tilde{N}}'=(N_2,\cdots,N_{L-1})$, $\mathbf{\tilde{a}}'=(a_2,\cdots,a_{L-1})$.

Then from (\ref{skewchain}), we obtain the superintegrability relation for the correlator
\begin{eqnarray}\label{skewtildJA}
	\widetilde{\langle J_{\lambda}(\mathbf{y}^{(1)}) \rangle}^{skew}_L=\sum_{\nu\subseteq\lambda} \prod_{l=1}^{L}\frac{\mathcal{J}^{(\beta)}_{\lambda}(N_l,a_l)}{\mathcal{J}^{(\beta)}_{\nu}(N_l,a_l)}\frac{J_{\lambda/\nu}(\mathbf{y})J_{\nu}(\mathbf{z})}{j_{\lambda}}.
\end{eqnarray}

The integral chain (\ref{skewchain}) may be expressed as
\begin{eqnarray}\label{schainop}
&&\bar{Z}_{chain}^{(\beta)}(\mathbf{x,y,z;a;N};L)\nonumber\\ &=&\hat{O}(\mathbf{\tilde{N}}\beta;\mathbf{x})\hat{O}^{-1}(\mathbf{\tilde{N}}
\beta;\mathbf{z})\hat{O}(\mathbf{a}+\Xi;\mathbf{x})\hat{O}^{-1}(\mathbf{a}+\Xi;\mathbf{z})\hat{O}^{-1}(\Xi;\mathbf{x})\hat{O}(\Xi;\mathbf{z})\nonumber\\
&&\cdot \exp\left(\beta\sum_{n=1}^{\infty} \frac{\hat{W}^{(1)}_{-n}(\xi_L;\mathbf{y})p_n(\mathbf{z})}{n}\right)\exp\left(\beta\sum_{n=1}^{\infty}
\frac{p_n(\mathbf{x})p_n(\mathbf{y})}{n}\right).
\end{eqnarray}

Comparing the character expansions (\ref{skewchain}) with (\ref{chain}), we have
\begin{subequations}
\begin{eqnarray}
	\bar{Z}^{(\beta)}_{chain}(\mathbf{x,y,0;a;N};L)&=&Z^{(\beta)}_{chain}(\mathbf{x,y;a;N};L)\nonumber\\
	&=&{_{2L+1}\mathcal{F}_L^{(\beta)}} (N\beta,\mathbf{N} \beta,\mathbf{a}+\Xi;\Xi;\mathbf{x;y}),\\
	\bar{Z}^{(\beta)}_{chain}(\mathbf{x,0,z;a;N};L)&=&Z^{(\beta)}_{chain}(\mathbf{x,z;a;N};0)\nonumber\\
	&=&{_{1}\mathcal{F}_0^{(\beta)}} (N\beta;\mathbf{x;z}).
\end{eqnarray}
\end{subequations}
Thus, we call (\ref{skewchain}) the skew version of the partition function (\ref{chain}).

Let us take $\mathbf{a}=\mathbf{0}$ and $\beta=1$ in (\ref{skewchain}), respectively, we have
\begin{eqnarray}
	&&\bar{Z}^{(\beta)}_{chain}(\mathbf{x,y,z;0;N};L)\nonumber\\
	&=&\frac{1}{I_{N_L}^{(0)}}\int_{\mathbf{T}_{N_L}}d\mathbf{y}^{(L)} \int_{\mathbb{R}_{+}^{N_L}} d\mathbf{x}^{(L)} \Delta^{2\beta} (\mathbf{x}^{(L)})\Delta^{2\beta} (\mathbf{y}^{(L)}){_0\mathcal{F}^{(\beta)}_0}(-\mathbf{x}^{(L)};\mathbf{y}^{(L)})\nonumber\\
	&&\times Z_{chain}^{(\beta)}(\mathbf{x},\mathbf{x}^{(L)}; \mathbf{0;\tilde{N}};L-1)\prod_{i=1}^{N_{L}}\prod_{j=1}^{N}(1-y_i^{(L)}y_j)^{-\beta}S(\mathbf{y}^{(L)},\mathbf{z};L;\beta)\nonumber\\
	&=&\exp\left(\beta\sum_{n=1}^{\infty}\frac{W^{(L)}_{-n}(\mathbf{\tilde{N}}\beta,\xi_L;\mathbf{y})p_n(\mathbf{z})}{n}\right)\exp\left(\beta\sum_{n=1}^{\infty}\frac{p_n(\mathbf{x})p_n(\mathbf{y})}{n}\right)\label{schain1},
\end{eqnarray}
and
\begin{eqnarray}
	&&\bar{Z}^{(1)}_{chain}(\mathbf{x,y,z;a;N};L)\nonumber\\
	&=&\frac{1}{I_{N_L}^{(a)}}\int_{\mathbf{T}_{N_L}}d\mathbf{y}^{(L)} \int_{\mathbb{R}_{+}^{N_L}} d\mathbf{x}^{(L)} \Delta^2(\mathbf{x}^{(L)})\Delta^2 (\mathbf{y}^{(L)}){_0\mathcal{F}^{(1)}_0}(-\mathbf{x}^{(L)};\mathbf{y}^{(L)})\prod_{j=1}^{N_L} (x_j^{(L)}y_j^{(L)})^{a_L}\nonumber\\
	&&Z_{chain}^{(1)}(\mathbf{x},\mathbf{x}^{(L)}; \mathbf{\tilde{a};\tilde{N}};L-1)\prod_{i=1}^{N_{L}}\prod_{j=1}^{N}(1-y_i^{(L)}y_j)^{-1}S(\mathbf{y}^{(L)},\mathbf{z};L;1)\nonumber\\
	&=&\exp\left(\sum_{n=1}^{\infty}\frac{\mathbb{W}^{(L)}_{-n}(\mathbf{a+N};\mathbf{y})p_n(\mathbf{z})}{n}\right)\exp\left(\sum_{n=1}^{\infty}\frac{p_n(\mathbf{x})p_n(\mathbf{y})}{n}\right).\label{schain2}
\end{eqnarray}
We see that the last lines of (\ref{schain1}) and (\ref{schain2}) reproduce the $W$-representations constructed in Refs.\cite{MirSkew,MirCom}.

\section{Conclusions}\label{Conclusion}
We have reinvestigated the CS-type models and analyzed the properties of corresponding CS operators. We have constructed the generalized CS operators $\tilde{\mathcal{H}}^{(X)}$ for circular,
Hermite, Laguerre, Jacobi and Bessel cases and established the generalized Lassalle-Nekrasov correspondence (\ref{LNC}). Based on the $\mathbf{SH}_N$ algebra, a family of $\hat W$-operators were constructed,
which can be used to represent the CS operators. Furthermore, to reveal the commutation relations of $\hat W$-operators, we constructed the $\hat O$-operator in terms of the generators
of Cartan subalgebra of $\mathbf{SH}_N$ algebra. We also presented two families of commutative $\hat W$-operators $\hat{W}_{n}^{(p)}$ and $\hat{W}_{-n}^{(p)}$  (\ref{Wops}),
which can be regarded as the Hamiltonians of certain many-body integrable systems.

We have analyzed the generalized hypergeometric functions. In terms of $\hat O$-operator (\ref{Odefine}), we presented the $O$-representations of the generalized hypergeometric functions (\ref{OrepF}). For the hypergeometric
functions with two set variables, the constraints (\ref{eigenWF}) were presented, where the constraint operators are given by the $\hat W$-operators. Moreover, some new properties associated with generalized hypergeometric functions and Hermite and Laguerre polynomials were presented, in addition to reproducing some well-known properties by new proofs. Not as the case of the hypergeometric functions with two
set variables, we derived the constraints (\ref{ConsF}) for hypergeometric functions with one set variables by the conjecture (\ref{WPKP}). Then from the several confirmed constraints, it follows that the multi-variable Jacobi,
Laguerre and Bessel polynomials with rectangular partition can be represented with generalized hypergeometric functions.

We have investigated the $\beta$-deformed integrals $Z^{(X)}\{\mathbf{t}\}$ (\ref{Zxt}) for $X\in\{C,H,L,J,B\}$, where the measures are associated with the corresponding ground state wave functions of circular,
Hermite, Laguerre, Jacobi and Bessel type CS models. We discussed the superintegrability relations for $Z^{(X)}$, $X\in\{H,L,J,B\}$.
Then by the generalized Laplace transformation of Jack polynomials, we constructed two integral chains. The integral chain $Z_{chain}^{(\beta)}(\mathbf{x,y;a;N};L)$ (\ref{chain}) is equal to the generalized hypergeometric function. Another one $\bar{Z}_{chain}^{(\beta)}(\mathbf{x,y,z;a;N};L)$ (\ref{skewchain}) is the skew version of (\ref{chain}). The superintegrability relations (\ref{tildJA}) and (\ref{skewtildJA}) have been derived.
Many well-known partition functions with $W$-representations turn out to be the special cases of our integral chains.
It is worthwhile to point out that due to the remarkable property of our integral chains, it is of interest to explore more applications of these integral chains for further research.

We have discussed the relations between CS-type systems and certain $\beta$-ensembles. It is known that such relations also emerge in Chern-Simons matrix models \cite{Tierz07},
$U(N|M)$-deformed matrix models \cite{Atai19} and ($q$,$t$)-deformed Selberg integrals \cite{Kaneko2}. It is interesting to construct more superintegrable models from other generalizations
of CS-type systems, especially the CS model with elliptic potential and ($q$,$t$)-deformed models.

\section *{Acknowledgments}

We are grateful to the referees of this paper for a series of valuable comments.
The authors dedicate this paper to professor Ke Wu for his eightieth birthday.
This work is supported by the National Natural Science Foundation of China (Nos. 12375004 and 12205368)
and the Fundamental Research Funds for the Central Universities, China (No. 2024ZKPYLX01).


\begin{thebibliography}{}
	
\bibitem{Calogero}
F. Calogero, Solution of a three problem in one dimension, J. Math. Phys. $\mathbf{10}$ (1969) 2197.
	
	
\bibitem{Sutherland}
B. Sutherland, Quantum many-body problem in one dimension: ground state, J. Math. Phys. $\mathbf{12}$ (1971) 246.
	

\bibitem{Langmann}
E. Langmann, Anyons and the elliptic Calogero-Sutherland model, Lett. Math. Phys. $\mathbf{54}$ (2000) 279, arXiv:math-ph/0007036.


\bibitem{Olshanetsky}
A.M. Olshanetsky and A.M. Perelomov, Quantum integrable systems related to Lie algebras, Phys. Rep. $\mathbf{94}$ (1983) 313.
	
	
\bibitem{Ruijsenaars86}
S.N.M. Ruijsenaars and H. Schneider, A new class of integrable systems and its relation to solitons, Ann. Phys. $\mathbf{170}$ (1986) 370.
	
	
\bibitem{Bernard}
D. Bernard, M. Gaudin, F.D.M. Haldane and V. Pasquier, Yang-Baxter equation in long-range interacting systems,  J. Phys. A: Math. Gen. $\mathbf{26}$ (1993) 5219, arXiv:hep-th/9301084.
	
	
\bibitem{Sergeev04}
A.N. Sergeev and A.P. Veselov, Deformed quantum Calogero-Moser problems and Lie superalgebras,
Commun. Math. Phys. $\mathbf{245}$ (2004) 249, arXiv:math-ph/0303025.
	
	
\bibitem{Sergeev09}
A.N. Sergeev and A.P. Veselov, Deformed Macdonald-Ruijsenaars operators and super Macdonald polynomials, Commun. Math. Phys. $\mathbf{288}$ (2009) 653, arXiv:0707.3129.
	
	
\bibitem{Ruijsenaars87}
S.N.M. Ruijsenaars, Complete integrability of relativistic Calogero-Moser systems and elliptic function identities, Commun. Math. Phys. $\mathbf{110}$ (1987) 191.
	
	
\bibitem{Lapointe}
L. Lapointe and L. Vinet, Exact operator solution of the Calogero-Sutherland Model, Commun. Math. Phys. $\mathbf{178}$ (1996) 425, arXiv:q-alg/9509003.
	
	
\bibitem{ForresterLog}
P.J. Forrester, \textit{Log-gases and random matrices}, Princeton University Press, Princeton, NJ, U.S.A. (2010).
	
	
\bibitem{Forrester1}
P.J. Forrester, Selberg correlation integrals and the $1/r^2$ quantum many-body system, Nucl. Phys. B $\mathbf{388}$ (1992) 671.


\bibitem{Forrester2}
P.J. Forrester, Addendum to `Selberg correlation integrals and the $1/r^2$ quantum many body system', Nucl. Phys. B $\mathbf{416}$ (1994) 377.


\bibitem{Forrester3}
P.J. Forrester, Recurrence equations for the computation of correlations in the $1/r^2$ quantum many-body system, J. Stat. Phys. $\mathbf{72}$ (1993) 39.
	
	
\bibitem{CMS}
J.F. van Diejen and L. Vinet, \textit{Calogero-Moser-Sutherland models}, Springer New York (2000).
	
	
\bibitem{BakerCMP}
T.H. Baker and P.J. Forrester, The Calogero-Sutherland model and generalized classical polynomials, Commun. Math. Phys. $\mathbf{188}$ (1997) 175, arXiv:solv-int/9608004.
	
	
\bibitem{Hallnas09}
M. Halln\"as, Multivariable Bessel polynomials related to the hyperbolic Sutherland model with external Morse
potential, Int. Math. Res. Not. $\mathbf{2009}$ (2009)1573, arXiv:0807.4740.
	
	
\bibitem{Hallnas10}
M. Halln\"as and E. Langmann, A unified construction of generalized classical polynomials associated with operators of Calogero-Sutherland type,
Constructive Approx. $\mathbf{31}$ (2010) 309, arXiv:math-ph/0703090.
	
	
\bibitem{Schiffmann}
O. Schiffmann and E. Vasserot,
Cherednik algebras, $W$ algebras and the equivariant cohomology of the moduli space of instantons on $\mathbb{A}^2$,
Publ. Math. Inst. Hautes \'Etudes Sci. $\mathbf{118}$ (2013) 213, arXiv:1202.2756.
	
	
\bibitem{Koranyi}
A. Kor{\'a}nyi, Hua-type integrals, hypergeometric functions and symmetric polynomials, in \textit{International symposium in memory of Hua Loo Keng}, Vol. 2, Springer (1998).
	
	
\bibitem{Macdonald13}
I.G. Macdonald, Hypergeometric functions I, arXiv:1309.4568.
	
	
\bibitem{Yan}
Z.M. Yan, A class of generalized hypergeometric functions in several variables, Can. J. Math. $\mathbf{44}$ (1992) 1317.
	
	
\bibitem{Kaneko}
J. Kaneko, Selberg integrals and hypergeometric functions associated with Jack polynomials, SIAM J. Math. Anal. $\mathbf{24}$ (1993) 1086.
	
	
\bibitem{Hikami}
K. Hikami and M. Wadati, Classical Calogero-Sutherland model and matrix model, J. Phys. Soc. Jpn. $\mathbf{62}$ (1993) 3857.
	
	
\bibitem{zhangwang}
C.H. Zhang and R. Wang, On $n$-algebraic structures in the Calogero model, Int. J. Mod. Phys. A 35 (2020) 2050137.
	
	
\bibitem{Lesage}
F. Lesage, V. Pasquier and D. Serban, Dynamical correlation functions in the Calogero-Sutherland model, Nucl. Phys. B 435 (1995) 585, arXiv:hep-th/9405008.
	
	
\bibitem{Awata95}
H. Awata, Y. Matsuo, S. Odake and J. Shiraishi, Collective field theory, Calogero-Sutherland model and generalized matrix models,
Phys. Lett. B $\mathbf{347}$ (1995) 49, arXiv:hep-th/9411053.
	
	
\bibitem{Awata96}
H. Awata, Y. Matsuo, S. Odake and J. Shiraishi, A note on Calogero-Sutherland model, $W_n$ singular vectors and generalized matrix models,
Soryushiron Kenkyu $\mathbf{91}$ (1995) A69, arXiv:hep-th/9503028.
	
	
\bibitem{Drachov}
Y. Drachov, Generalized $\widetilde{W}$ algebras, Eur. Phys. J. C $\mathbf{84}$ (2024) 1078, arXiv:2406.13624
	
	
\bibitem{Morozov09}
A. Morozov and Sh. Shakirov, Generation of matrix models by $\hat{W}$-operators, JHEP $\mathbf{04}$ (2009) 064, arXiv:0902.2627.
	
	
\bibitem{Cassia}
L. Cassia, R. Lodin and M. Zabzine, On matrix models and their $q$-deformations, JHEP $\mathbf{10}$ (2020) 126, arXiv:2007.10354.
	
	
\bibitem{MirGKM}
A. Mironov, V. Mishnyakov and A. Morozov, Non-Abelian $W$-representation for GKM, Phys. Lett. B $\mathbf{823}$ (2021) 136721, arXiv:2107.02210.
	
	
\bibitem{MirSum}
A. Mironov and A. Morozov, Superintegrability summary, Phys. Lett. B $\mathbf{835}$ (2022) 137573, arXiv:2201.12917.
	
	
\bibitem{Rui22}
R. Wang, F. Liu, C.H. Zhang and W.Z. Zhao,
Superintegrability for ($\beta$-deformed) partition function hierarchies with $W$-representations,
Eur. Phys. J. C $\mathbf{82}$ (2022) 902, arXiv:2206.13038.
	
	
\bibitem{MirSkew}
A. Mironov, V. Mishnyakov, A. Morozov, A. Popolitov and W.Z. Zhao, On KP-integrable skew Hurwitz $\tau$-functions and their $\beta$-deformations,
Phys. Lett. B \textbf{839} (2023) 137805, arXiv:2301.11877.
	
	
\bibitem{MirInter}
A. Mironov, V. Mishnyakov, A. Morozov, A. Popolitov, R. Wang and W.Z. Zhao, Interpolating matrix models for WLZZ series, Eur. Phys. J. C $\mathbf{83}$ (2023) 377, arXiv:2301.04107.
	
	
\bibitem{Oreshina03}
A. Mironov, A. Oreshina and A. Popolitov, Two $\beta$-ensemble realization of $\beta$-deformed WLZZ models, Eur. Phys. J. C \textbf{84} (2024) 705, arXiv:2403.05965.
	
	
\bibitem{Oreshina04}
A. Mironov, A. Oreshina and A. Popolitov,
$\beta$-WLZZ models from $\beta$-ensemble integrals directly, JETP Lett. $\mathbf{120}$ (2024) 66, arXiv:2404.18843.
	
	
\bibitem{Eynard}
M. Berg\`{e}re and B. Eynard, Some properties of angular integrals, J. Phys. $\mathbf{A}$: Math. Theor. $\mathbf{42}$ (2009) 265201, arXiv:0805.4482.
	
	
\bibitem{MirMany}
A. Mironov and A. Morozov, Many-body integrable systems implied by WLZZ models, Phys. Lett. B $\mathbf{842}$ (2023) 137964, arXiv:2303.05273.
	
	
\bibitem{MirCom}
A. Mironov, V. Mishnyakov, A. Morozov and A. Popolitov, Commutative families in $W_{\infty}$, integrable many-body systems and hypergeometric $\tau$-functions,
JHEP $\mathbf{09}$ (2023) 065, arXiv:2306.06623.
	
	
\bibitem{Alexandrov14}
A. Alexandrov, A. Mironov, A. Morozov and S. Natanzon,
On KP-integrable Hurwitz functions, JHEP $\mathbf{11}$ (2014) 080, arXiv:1405.1395.
	
	
\bibitem{Alexandrov24}
A. Alexandrov, On $W$-operators and superintegrability for dessins d'enfant, Eur. Phys. J. C $\mathbf{83}$ (2023) 147, arXiv:2212.10952.
	
	
\bibitem{Chapuy1}
G. Chapuy and M. Do{\l}{\c{e}}ga, Non-orientable branched coverings, $b$-Hurwitz numbers, and
positivity for multiparametric Jack expansions, Adv. Math. $\mathbf{409}$ (2022) 108645, arXiv:2004.07824.
	
	
\bibitem{Chapuy2}
V. Bonzom, G. Chapuy and M. Do{\l}{\c{e}}ga, b-Monotone Hurwitz Numbers: Virasoro Constraints, BKP Hierarchy, and $O(N)$-BGW Integral,
Int. Math. Res. Not. $\mathbf{2023}$ (2023) 12172, arXiv:2109.01499.
	
	
\bibitem{Stanley89}
R. Stanley, Some combinatorial properties of Jack symmetric functions, Adv. Math. $\mathbf{77}$ (1989) 76.
	
	
\bibitem{Macdonald}
I.G. Macdonald, \textit{Symmetric functions and Hall polynomials,}  2nd ed., Oxford University Press, New York, U.S.A. (1995).
	
	
\bibitem{Lassalle1}
M. Lassalle, Polyn\^{o}mes de Hermite g\'{e}n\'{e}ralis\'{e}s, C. R. Acad. Sci. Paris S\'{e}r. I Math. $\mathbf{312}$ (1991) 579.


\bibitem{Lassalle2}
M. Lassalle, Polyn\^{o}mes de Laguerre g\'{e}n\'{e}ralis\'{e}s, C. R. Acad. Sci. Paris S\'{e}r. I Math. $\mathbf{312}$ (1991) 725.

\bibitem{Lassalle3}
M. Lassalle, Polyn\^{o}mes de Jacobi g\'{e}n\'{e}ralis\'{e}s, C. R. Acad. Sci. Paris S\'{e}r. I Math. $\mathbf{312}$ (1991) 425.

	
\bibitem{Sogo}
K. Sogo, A simple derivation of multivariable Hermite and Legendre polynomials, J. Phys. Soc. Jpn. $\mathbf{65}$ (1996) 3097.
	
	
\bibitem{Desrosiers}
P. Desrosiers and M. Halln\"{a}s, Hermite and Laguerre symmetric functions associated with operators of Calogero-Moser-Sutherland type,
SIGMA $\mathbf{8}$ (2012) 049, arXiv:1103.4593.
	
	
\bibitem{Nekrasov}
N. Nekrasov, On a duality in Calogero-Moser-Sutherland systems, arXiv:hep-th/9707111.


\bibitem{Cherednik}
I. Cherednik, \textit{Double Affine Hecke Algebras}, London Mathematical Society Lecture Note Series, vol. 319, Cambridge
University Press (2005).
	
	
\bibitem{Fan}
F. Liu, R. Wang, J. Yang and W.Z. Zhao, Generalized $\beta$ and $(q,t)$-deformed partition functions with $W$-representations and Nekrasov partition functions,
Eur. Phys. J. C $\mathbf{84}$ (2024) 756, arXiv:2405.11970.
	
	
\bibitem{Erdelyi}
A. Erd\'elyi, et al., \textit{Higher Transcendental Functions}, Vol. 1. McGraw-Hill, New York, U.S.A. (1953).
	
	
\bibitem{MirBGW}
A. Mironov, A. Morozov and S. Shakirov, Brezin-Gross-Witten model as ``pure gauge" limit of Selberg integrals, JHEP $\mathbf{03}$ (2011) 102, arXiv:1011.3481.
	
	
\bibitem{Heckman}
G.J. Heckman, An elementary approach to the hypergeometric shift operators of Opdam, Invent. Math. $\mathbf{103}$ (1991) 341.


\bibitem{Diejen}
J.F. van Diejen, Properties of some families of hypergeometric orthogonal polynomials in several variables, Trans. Amer. Math. Soc. $\mathbf{351}$ (1999) 233, arXiv:q-alg/9604004.
	

\bibitem{Dumitriu}
I. Dumitriu and A. Edelman, Matrix models for beta ensembles,  J. Math. Phys. $\mathbf{43}$ (2002) 5830, arXiv:math-ph/0206043.

	
\bibitem{Kadell}
K.W.J. Kadell, The Selberg-Jack polynomials, Adv. Math. $\mathbf{130}$ (1997) 33.
	

\bibitem{Orlov}
A. Yu. Orlov, Tau-functions and matrix integrals, arXiv:math-ph/0210012.


\bibitem{Alexandrovchain}
A. Alexandrov, Matrix models for the nested hypergeometric tau-functions, Commun. Num. Theor. Phys. \textbf{19} (2025) 241, arXiv:2304.03051.


\bibitem{Tierz07}
Y. Dolivet and M. Tierz, Chern-Simons matrix models and Stieltjes-Wigert polynomials, J. Math. Phys. $\mathbf{48}$ (2007) 023507, arXiv:hep-th/0609167.
	
	
\bibitem{Atai19}
F. Atai, M. Halln\"as and E. Langmann, Orthogonality of super-Jack polynomials and a Hilbert space
interpretation of deformed Calogero-Moser-Sutherland operators, Bull. Lond. Math. Soc. $\mathbf{51}$ (2019) 353, arXiv:1802.02016.
	
	
\bibitem{Kaneko2}
J. Kaneko, $q$-Selberg integrals and Macdonald polynomials, Ann. Sci. \'{E}cole Norm. Sup. $\mathbf{29}$ (1996) 583.
	
\end{thebibliography}
\end{document}